\documentclass[lettersize,journal]{IEEEtran}
\usepackage{amsmath,amsfonts, amsthm, amssymb}
\usepackage{algorithm}
\usepackage{array}
\usepackage[caption=false,font=normalsize,labelfont=sf,textfont=sf]{subfig}
\usepackage{textcomp}
\usepackage{stfloats}
\usepackage{url}
\usepackage{verbatim}
\usepackage{graphicx}
\usepackage{cite}

\usepackage{algorithm}
\usepackage{arevmath}   
\usepackage{algpseudocode}
\usepackage{authblk}
\usepackage{xcolor}
\usepackage{graphicx}
\usepackage[most]{tcolorbox}
\usepackage{amsthm}
\newtheorem{assumption}{Assumption}
\newtheorem{theorem}{Theorem}
\usepackage[most]{tcolorbox}
\usepackage{listings}
\usepackage{xcolor} 
\usepackage{adjustbox}
\usepackage{multirow}

\definecolor{dkgreen}{rgb}{0,0.6,0}
\definecolor{mauve}{rgb}{0.58,0,0.82}
\definecolor{darkgreen}{RGB}{0, 128, 0}

\newtcolorbox{definitiongray}[1]{%
  colback=gray!5!white,
  colframe=gray!75!black,
  fonttitle=\bfseries,
  title={#1},
  arc=1mm, 
  before=\vspace{3mm},
  after=\vspace{3mm}
}

\begin{document}

\title{Efficient Fault Localization in a Cloud Stack Using
End-to-End Application Service Topology}

\author[1]{Dhanya R Mathews}
\author[2]{Mudit Verma}
\author[2]{Pooja Aggarwal}
\author[1]{J.Lakshmi}
\affil[1]{Indian Institute of Science, Bangalore, India}
\affil[2]{IBM Research, India}



\maketitle

\begin{abstract}
Cloud application services are distributed and have components across the stack working together to deliver the end-user experience. The widespread adoption of microservice architecture exacerbates failure management due to the increased number of service components. To be effective, strategies to enhance application service resilience need to be autonomous and developed at the service granularity, considering its end-to-end components. However, the massive amount of observability data generated by all these components across the service stack poses a significant challenge in reacting to anomalies and restoring the service quality in real time. 
Towards this end, the article presents the Topology-Aware Root Cause Detection (TA-RCD) algorithm, which leverages the end-to-end application service topology to accurately pinpoint root causes and facilitate swift remediation of cloud service anomalies. Additionally, because effective root-cause localization depends on identifying the most informative observability signals across the cloud service stack, this article presents a topology-guided metric selection strategy that selects relevant metrics from end-to-end layers to enable efficient, explainable, and accurate anomaly remediation. 
TA-RCD is comprehensively evaluated on the public RCAEval benchmark and a controlled in-house testbed. Experimental results show that the proposed approach consistently achieves higher average Top-3 and Top-5 localization accuracy than state-of-the-art baselines, improving AC@5 by at least 2$\times$ and AC@3 by at least 4$\times$ over standard RCD in the controlled testbed. On the public RCAEval benchmark, TA-RCD further improves metric-level AC@5 by 18\% and Avg@5 by 19\% over the strongest baseline.
\end{abstract}


\renewcommand{\thefootnote}{\fnsymbol{footnote}}
\footnotetext[1]{This work has been submitted to the IEEE for possible publication. Copyright may be transferred without notice, after which this version may no longer be accessible.}

\section{Introduction}


Cloud computing delivers geographically distributed infrastructure and software as a utility via service-oriented architectures \cite{cloud_computing}. To guarantee end-user experience, providers must strictly adhere to Service Level Objectives (SLOs) defined in Service Level Agreements (SLAs). Consequently, SLO violations serve as primary indicators of application service failures, typically triggered by anomalies within service instances or their dependencies. An anomaly results from a fault, error, or failure in the service instances or their dependent component(s). These anomalies can originate at any architectural layer, propagate horizontally or vertically across components, and ultimately degrade user-perceived service quality \cite{drm_ucc}. 

Manually addressing anomalies at the cloud scale is complex, given the number of components across the logical layers involved in delivering the service and the volume of data generated at this scale. A cloud application service is identified as resilient if it can deliver the guaranteed SLOs even during anomalies \cite{res_defn}. Hence,  autonomously identifying and remediating anomalies is inevitable. Cloud setups rely on highly autonomous service layers for managing, provisioning, and monitoring applications. However, the capability for a service to demonstrate competence in resisting various types of anomalies or to re-establish normal operations in the minimum possible time can be built only if autonomous strategies to ensure its resilience are devised at the granularity of the cloud application service \cite{drm1}.

Autonomous efforts to ensure the resilience of a cloud application service require a precise understanding of what caused the anomaly and where it occurred across the end-to-end application service stack. Such an understanding requires the consideration of end-to-end service components across different logical layers \cite{drm1}. Most efforts to manage application service anomalies follow a black-box approach \cite{rcd} and are opaque to the causes arising from lower layers of the stack. In microservice architectures, it is often challenging to assess how individual components affect the overall service delivery \cite{deathstar}. Enhancing cloud service resilience, therefore, depends on comprehending the roles of components across different layers, bridging the semantic gap between architectural levels, and effectively addressing identified anomalies. Moreover, effective failure management at cloud scale must be explainable, which, in turn, relies on accurate observability of failure causes across the service stack.
 
Motivated by the need for explainable and effective failure management at cloud scale, this work observes that anomaly localization accuracy improves with end-to-end service topology awareness. Towards this, the article proposes Topology-Aware Root Cause Detection (TA-RCD), a root-cause localization approach that leverages application service topology to guide causal analysis and improve fault localization accuracy. To address the large volume of observability data collected  from across end-to-end service layers, the article also adopts a systematic, reasoning-based metric selection strategy to identify actionable cross-layer metrics. Experimental results show that TA-RCD localizes anomaly root causes more effectively than the evaluated baselines while supporting more explainable remediation decisions.

To summarize, the key contributions of this article are:
\begin{itemize}
\item Present a systematic end-to-end metric selection strategy to identify actionable observability signals from across the cloud stack, including service, container, and infrastructure layers, for detecting cloud service anomalies.
\item Propose Topology-Aware Root Cause Detection (TA-RCD), a topology-guided root-cause localization approach that leverages end-to-end application service topology to improve the accuracy, efficiency, and explainability of root-cause analysis.
\item Establish the effectiveness of TA-RCD through empirical evaluation against representative root-cause localization baselines across multiple benchmark datasets and fault scenarios.
\end{itemize}

The remainder of this article is organized as follows. Section \ref{background} motivates the work. The related works are summarised in Section \ref{related_work}. Section \ref{met_selection} explains the metric selection strategy, and Section \ref{rcd} details the root cause localization technique. Section \ref{eval} presents the evaluation results, and Section \ref{conclusion} concludes the paper.

\section{Background and Motivation} \label{background}
Cloud application services comprise interdependent components across three logical layers: Infrastructure, Middleware, and Application Services. The Infrastructure layer provides the physical and virtual data center resources required for computing, storage, and networking. The Middleware layer abstracts these hardware resources to manage service instances and infrastructure allocation, while the Application Services layer encompasses the specific service instances of the microservices application. Within this architecture, the serviceability of any given component depends directly on the performance and availability of its connected components within the same or underlying layers.

In a cloud service environment, the available observability data is identified as the golden triangle \cite{dso} and comprises metrics, event logs, and traces. This work uses metrics available across layers of the cloud stack for observability to localize failures in real time. A \textit{metric} is an actual value collected from a component \cite{metric_defn} that externalizes the system's internal state and helps detect anomaly scenarios and SLO variations \cite{metric_expln}. The monitoring module collects the metric values at predefined time intervals. The vector of metrics collected at a given timestamp is called a monitoring observation \cite{metric_defn}. The scale of available metrics across cloud service architecture layers makes end-to-end observability a formidable challenge.
To exemplify the magnitude of metrics across the cloud stack of real-world microservice applications, Netflix exposes approximately 2,000,000 metrics, Uber exposes around 500,000,000 metrics, and Quantcast around 2,000,000 metrics \cite{sieve}. Additionally, at any given time, a cloud service can be in a normal, pre-failure, or failed state \cite{ubl}. Identifying the anomaly in a pre-failure state will increase the lead time and, hence, the time to act on the failure.

Given the distributed nature of cloud services, failures in individual components can have widespread impact across the service. In such a scenario, looking through all available observability data across these distributed components is ineffective for two major reasons. Primarily, large amounts of observability data can be misleading in identifying the real cause of failure due to confounding correlations. Secondly, real-time root-cause remediation depends on the precise identification of relevant telemetry data. Proper selection of metrics can thus narrow the search space while accounting for all components involved in service delivery, thereby enabling an end-to-end approach.

 Furthermore, treating cloud systems as black boxes without considering component and layer semantics reduces diagnostic explainability and impedes efficient failure management. Because autonomous failure management requires clear causal justification, leveraging end-to-end service topology is critical to effectively capturing dependencies between service components. In microservice architectures, direct edges within service or resource graphs indicate a higher probability of causal influence due to layered, API-based interactions. Incorporating this structural knowledge enhances both the efficiency of root-cause analysis and the explainability of remediation strategies. Therefore, root-cause localization must consider the structural interconnectivity and semantics of service-delivery components. To achieve this, this work utilizes an end-to-end cloud service topology \cite{drm_topo} combined with cross-stack metric selection to enable effective, explainable localization of SLO anomalies.

\section{Related Work} \label{related_work}
This work surveys prior research to investigate the rationale behind multi-layer cloud metric selection and the methodologies used for root-cause localization.

\textbf{Metrics across cloud service architecture layers: }Monitoring techniques recognized by cloud service providers primarily gravitate around three methodologies: Google's Golden Signals (Latency, Traffic, Errors, Saturation) \cite{sre_golden_sig}, Brendon Gregg's Utilization, Saturation, and Error (USE) method for kernel and resource bottleneck identification \cite{bgregg}, and Weaveworks' Request Rate, Errors, and Duration (RED) metrics for microservices \cite{red}.

In the domain of reactive failure management, literature heavily leverages these utilization and saturation signals for localization and anomaly diagnosis. Container-level resource metrics (CPU, memory, load) are tracked by MicroCause \cite{microcause}, RCD \cite{rcd}, and \cite{ad_container}. This resource-driven approach is extended by \cite{frlmpfg}, \cite{perf_diag_micro}, and MicroRCA \cite{microrca}, which combine host/container metrics with application-layer response times. To localize faults, broader system-level metrics are employed in \cite{fd4c}, FChain \cite{fchain}, \cite{loud}, \cite{wang2023hierarchical}, and \cite{frlmpfg}, while AutoMAP \cite{automap} incorporates power consumption and service availability. Conversely, some frameworks prioritize execution performance over resource consumption; CloudRanger \cite{cloudranger} isolates latency as the primary indicator of abnormality, and CauseInfer \cite{causeinfer} utilizes a unified TCP request latency metric. For multi-tier and three-tier architectures, root-cause analysis relies on VM-level throughput \cite{rca_pc} or targeted database and CPU metrics \cite{cid_smallmet}.

Proactive failure prediction techniques similarly depend on these underlying utilization metrics. Frameworks such as HORA \cite{hora} and \cite{hora1} monitor CPU, memory, load averages, and JVM heap/swap spaces alongside response times to forecast system degradation. Similarly, data-center and VM-level predictors, including \cite{fail_pred1}, PREPARE \cite{prepare}, and UBL \cite{ubl}, rely on OS or hypervisor-level resource traffic, whereas \cite{lep} integrates resource data with configuration and instance interaction metrics for latent error prediction.

Despite the breadth of existing literature, the rationale for selecting specific metrics remains largely unaddressed. Because anomaly explainability is strictly bounded by system observability, a principled metric selection strategy is critical to ensure unambiguous, consistent, and complete diagnostic evidence \cite{fl_backgnd}. Explaining any failure management mechanism requires a comprehensive understanding of monitored metrics, their structural significance, and their respective architectural layers. To address this gap, the proposed framework incorporates a metric selection strategy grounded in a systematic analysis of available telemetry and application-layer semantics.

\textbf{Root cause localization: } Most root-cause localization frameworks deploy a combination of causality and inference algorithms to isolate anomalies \cite{microcause, cloudranger, caus_mining, microrca, root_cause_detect_soa, rca_pc, automap}, evaluating efficiency via metrics such as Accuracy (AC) \cite{microcause, cloudranger}, Recall \cite{rcd}, Average Accuracy (Avg), Precision (PR) \cite{microrca, root_cause_detect_soa, automap}, and Mean Average Precision (MAP) \cite{microrca, root_cause_detect_soa}. Graph-based approaches are predominant: MicroCause \cite{microcause} pairs Path Condition Time Series (PCTS) with Temporal Cause-Oriented Random Walk (TCORW); CloudRanger \cite{cloudranger} and \cite{caus_mining} use variants of the PC algorithm combined with random walks; and MicroRCA \cite{microrca} utilizes an attributed anomalous subgraph with Personalized PageRank. Similarly, MonitorRank \cite{root_cause_detect_soa}, \cite{rca_pc}, and AutoMAP \cite{automap} employ random walks over dependency or similarity graphs.

Alternative structural and statistical models are also prevalent. BARO \cite{baro} integrates multivariate Bayesian online change point detection with a robust distribution-based scoring mechanism to pinpoint failure onset and rank root causes. CausalRCA \cite{causalrca} uses a gradient-based structure learning method to generate weighted causal graphs and subsequently infers localized root-cause metrics. TORAI introduces an unsupervised approach that clusters microservices based on multi-source telemetry anomaly severity to rank root causes without relying on an explicit service call graph \cite{torai}. CIRCA \cite{causal_infer} uses architectural knowledge to construct metric graphs for regression hypothesis testing, while Sage \cite{sage} models RPC-level traces using a Causal Bayesian Network and a graphical variational autoencoder. Outlier and distribution analyses are leveraged in \cite{metric_rank} via anomaly pattern classification, in \cite{perf_diag_dl} through autoencoder reconstruction errors, and in $\epsilon$-Diagnosis \cite{ediag} using the coefficient of variation (COV). Furthermore, trace- and log-based causal graphs are constructed by MRCA \cite{mrca}, self-supervised neural Granger causal discovery is applied in RUN \cite{run}, and the constrained $\psi$-PC algorithm is implemented in RCD \cite{rcd}

While many existing studies build large metric-level anomaly propagation graphs, such graphs can be computationally expensive and less interpretable than compact structural service topologies. To reduce overhead and improve explainability, this study uses paths derived from the physical and logical service topology for root-cause localization. Accordingly, BARO, RCD, and CIRCA are selected as metric-based baselines for comparison, while RUN and CausalRCA are excluded because they did not complete within the experimental timeline.

The subsequent sections present the proposed metric selection framework and delineate the TA-RCD algorithm.

\section{Metric Selection Strategy for Effective Root-Cause Localization} \label{met_selection}
The intricate dependencies and constant evolution of cloud service layers allow component-level anomalies to rapidly propagate through the stack, complicating failure management. Achieving explainability requires understanding both the component changes and how they manifest as service anomalies. As predefining all failure scenarios in such dynamic environments is unscalable, failure management mechanisms must adaptively trace cross-layer changes. Consequently, identifying these structural variations, mapping them to relevant telemetry signals, and utilizing the resulting metrics is critical for effectively monitoring SLOs.
From a root-cause localization standpoint, this work analyzed the following questions:
\begin{itemize}
\item What changes across the layers in a cloud stack?
\item What metrics are needed to reason about the changes?
\end{itemize}

\vspace{-0.7em}
\subsection{The Changes Leading to Performance Anomalies Across the Cloud Service Stack}
As users access the service instance APIs, the demand changes in the service instances layer. This change in demand translates to changes in demand for components in the underlying layers. 
The finite capabilities of the components in the underlying layers, in terms of total capacity or the limits imposed on compute, storage, and networking resources, are affected by load dynamics and/or performance interferences (other processes contending for the resources) in each of these components. Such changes in the cloud stack lead to anomalies in servicing the demand. Consequently, both \textit{Demand} and \textit{Serviceableness} vary dynamically across the architectural layers of the cloud stack.
\vspace{-0.7em}
\subsection{Monitoring the Changes Across the Cloud Service Stack}
Monitoring demand and serviceableness signals across all architectural layers is critical to isolating bottlenecks and explaining cloud service anomalies. Effectively capturing these signals requires identifying specific telemetry metrics from each layer involved in service delivery. For readability, this section uses the term \textit{node} to refer to either a service instance or an underlying infrastructure component along the service delivery path.

A node’s \textbf{activity} directly reflects its operational demand, with relevant metrics determined by its specific architectural layer. At the service instances layer, activity typically corresponds to workload, represented by active users or user requests. Conversely, at the physical or virtual machine layers, activity is captured via resource-specific indicators: CPU operations, memory reads/writes, disk I/O, and network packet or byte transmissions.

Similarly, a node’s serviceableness is determined by whether it is in a \textbf{state} to deliver its expected functionality. To quantify this across the architecture, this work monitors RED (Request Rate, Error, Duration) metrics at the service instances layer and USE (Utilization, Saturation, and Error) metrics across the underlying container, VM, and physical machine layers. Under this framework, a component's state is inherently tied to its activity, making activity metrics highly effective for explaining state changes. For example, at the microservice level, user request volume dictates activity while request duration defines the state. At the infrastructure level, increased CPU demand manifests as elevated utilization, while other resource states are similarly derived from their physical characteristics. Guided by a manual exploration of the Prometheus ecosystem, these distinct activity and state signals were mapped to specific, measurable metrics for each architectural tier (Table \ref{table_metrics}). Finally, mathematical aggregators, including \textit{rate} and \textit{histogram\_quantile}, are employed to transform these raw metrics into meaningful performance inferences.

\begin{table*}[]
\caption{Metrics Selected from Across the Cloud Stack for Monitoring}
\scriptsize
\centering
\begin{tabular}{|l|l|l|l|}
\hline
\textbf{Layer} &
  \textbf{Component} &
  \textbf{Metric Class} &
  \textbf{Metrics for Monitoring} \\ \hline
\multirow{2}{*}{Service Instance Layer} &
  \multirow{2}{*}{Microservice instance} &
  State &
  \begin{tabular}[c]{@{}l@{}}Request Rate, Latency, Error rate, \\Any microservice specific metric of interest\end{tabular} \\ \cline{3-4} 
 &
   &
  Activity &
  \begin{tabular}[c]{@{}l@{}}Number of user request or number of active users\end{tabular} \\ \hline
\multirow{8}{*}{Container Layer} &
  \multirow{2}{*}{Container CPU} &
  State &
  \begin{tabular}[c]{@{}l@{}}Usage/Utilization, Saturation (Container Throttling, \\Run-queue wait seconds)\end{tabular} \\ \cline{3-4} 
 &
   &
  Activity &
  \begin{tabular}[c]{@{}l@{}}Container CPU context switches, Container CPU Migrations\end{tabular} \\ \cline{2-4} 
 &
  \multirow{2}{*}{Container Memory} &
  State &
  \begin{tabular}[c]{@{}l@{}}Usage/Utilization, Swap usage, Saturation, OOM Failures\end{tabular} \\ \cline{3-4} 
 &
   &
  Activity &
  \begin{tabular}[c]{@{}l@{}}(Cache misses/Cache references), Page faults (Major and Minor)\end{tabular} \\ \cline{2-4} 
 &
  \multirow{2}{*}{Container filesystem and disk} &
  State &
  \begin{tabular}[c]{@{}l@{}}Page cache usage, Seconds spent doing I/O (\%util in iostat),\\ Number of queued operations, Avg. time to serve reads and writes\end{tabular} \\ \cline{3-4} 
 &
   &
  Activity &
  I/O throughput, I/O rate \\ \cline{2-4} 
 &
  \multirow{2}{*}{Container network} &
  State &
  \begin{tabular}[c]{@{}l@{}}Total bytes transmitted and received,
  Packets dropped, \\
  Transmit and Received Errors\end{tabular} \\ \cline{3-4} 
 &
   &
  Activity &
  \begin{tabular}[c]{@{}l@{}}Transmit and Receive throughput,
  Transmit and Receive packet rate\end{tabular} \\ \hline
\multirow{8}{*}{\begin{tabular}[c]{@{}l@{}}Infrastructure Layer\\(Virtual and Physical Machines)\end{tabular}} &
  \multirow{2}{*}{CPU} &
  State &
  \begin{tabular}[c]{@{}l@{}}Usage/Utilization, Saturation (Load trend, Run-queue wait seconds)\end{tabular} \\ \cline{3-4} 
 &
   &
  Activity &
  \begin{tabular}[c]{@{}l@{}}CPU context switches, CPU Migrations\end{tabular} \\ \cline{2-4} 
 &
  \multirow{2}{*}{Memory} &
  State &
  \begin{tabular}[c]{@{}l@{}}Usage/Utilization, Swap usage, Saturation (pswpin+pswpout)\end{tabular} \\ \cline{3-4} 
 &
   &
  Activity &
  \begin{tabular}[c]{@{}l@{}}(LL Cache misses/LL Cache references), Page faults (Major and Minor)\end{tabular} \\ \cline{2-4} 
 &
  \multirow{2}{*}{Filesystem and Disk} &
  State &
  \begin{tabular}[c]{@{}l@{}}Filesystem utilization,
  Inode Utilization, \\Seconds spent doing I/O (\%util in iostat),\\ Average queue length, Avg. wait time for reads and writes,\\ Filesystem device error\end{tabular} \\ \cline{3-4} 
 &
   &
  Activity &
  \begin{tabular}[c]{@{}l@{}}I/O throughput, I/O rate, CPU seconds in IO wait\end{tabular} \\ \cline{2-4} 
 &
  \multirow{2}{*}{Network} &
  State &
  \begin{tabular}[c]{@{}l@{}}Total bytes transmitted and received,
  Transmit queue length,\\
  Packets dropped, 
  Transmit and Received Errors\end{tabular} \\ \cline{3-4} 
 &
   &
  Activity &
  \begin{tabular}[c]{@{}l@{}}Transmit and Receive throughput,
  Transmit and Receive packet rate\end{tabular} \\ \hline
\end{tabular}
\label{table_metrics}
\end{table*}
\vspace{-0.5cm}
\subsection{Adequacy of the Selected Metrics for Anomaly Localization}
Both the USE \cite{bgregg} and RED \cite{red} methodologies are sufficient for identifying architectural anomalies based on layer semantics, as any performance degradation inherently manifests through their core metrics. Consequently, they can isolate anomalous components with high confidence during root-cause localization. However, incorporating activity metrics alongside them extends this localization to include driving workload behavior, aligning with industry-standard practices that prioritize identifying the most probable root causes. This approach provides much greater explainability, as demonstrated by the experiments presented in this paper.
\vspace{-0.5cm}
\subsection{Topology Discovery Tool and Metric Endpoint}
\label{metric_endpoint}
For local deployments, the cross-layer dynamic service topology discovery tool proposed in \cite{drm_topo} is used to identify the end-to-end service components. The tool is selected because it does not require any changes to the microservices, identifies the service topology from the deployment files, and tracks environment changes to update the generated topology information. 
The tool performs end-to-end discovery by traversing the entire service stack. At the service instance layer, it utilizes \textit{Istio Kiali} to map microservices and their interactions. It then queries \textit{Kubernetes APIs} to link these services to their respective containers, pods, and virtual nodes, capturing granular metadata for vCPU, memory, and networking. Finally, via \textit{libvirt management APIs}, it identifies the underlying physical resources. By correlating these layers, the tool establishes a complete vertical mapping from service instance to physical node, providing a holistic view of the service delivery topology.
This work considers the identified end-to-end components for metric selection and root-cause localization. However, for the RCAEval datasets used in this study, the topology information is derived from metadata files containing service interconnections and service-to-infrastructure mappings.
\vspace{-0.7em}
\subsection{Reduction in Metrics Dimension}
The total number of state and activity metrics yielded by the selection strategy for each locally deployed component is summarized by architectural layer in Table \ref{selected_metrics}. As can be seen, the number of selected metrics is significantly lower than the total available in each layer, though more than just the state metrics. The following sections demonstrate the impact of an informed selection of metrics across the cloud stack on the accuracy and efficiency of fault localization techniques.
\begin{table}[!h]
\caption{Number of Metrics Selected by the Proposed Strategy}
\centering
\begin{tabular}{|l|l|l|}
\hline
\textbf{\begin{tabular}[c]{@{}l@{}}Cloud Service \\ Architecture Layer\end{tabular}} &
  \textbf{\begin{tabular}[c]{@{}l@{}}Available \\ metrics\end{tabular}} &
  \textbf{\begin{tabular}[c]{@{}l@{}}Selected \\ metrics\end{tabular}} \\ \hline
\textit{\begin{tabular}[c]{@{}l@{}}Service Instances \\ Layer\end{tabular}} &
  \begin{tabular}[c]{@{}l@{}}5-58 (depending\\ on microservice)\end{tabular} &
  \begin{tabular}[c]{@{}l@{}}5 (per \\ microservice)\end{tabular} \\ \hline
\textit{Container Layer} &
  110 &
  29 \\ \hline
\textit{\begin{tabular}[c]{@{}l@{}}Infrastructure Layer \\ (Both Virtual Machine \\ and Physical Machine\\ Layer)\end{tabular}} &
  \begin{tabular}[c]{@{}l@{}}292 (at VM) \\ 336 (at PM)\end{tabular} &
  \begin{tabular}[c]{@{}l@{}}34 (each for VM \\ and PM)\end{tabular} \\ \hline
\end{tabular}
\label{selected_metrics}
\end{table}

\section{Root Cause Localization for Efficient Autonomous Failure Management}
\label{rcd}
Accurate root-cause localization is crucial for executing effective remediation strategies within autonomous failure management frameworks. As modern cloud applications have stringent latency constraints, this localization must be precise and occur in real time. The subsequent sections outline the baseline algorithm (RCD) and describe the enhancements implemented to optimize root-cause localization efficiency.
\vspace{-1.0em}
\subsection{RCD: The Baseline algorithm}
\label{baseline}

The baseline algorithm used in this work, RCD \cite{rcd}, is a localized, hierarchical causal discovery framework. It is grounded in the key insight that a failure alters the underlying generative mechanism of the affected node; consequently, identifying the root cause does not require reconstructing the global causal graph. RCD is highly relevant as a baseline amid alternative localization approaches, owing to its intervention-based causal inference, its efficient scaling without full-graph overhead, and its direct execution over telemetry metrics.

RCD models failures as interventions on a root-cause node, accepting a normal time-series dataset $D = \{ M(1, 1), \dots, M(n, t-1) \}$ and an anomalous dataset $D^* = \{ M(1, t), \dots, M(n, E) \}$ as input. Here, $M(i,k)$ represents a set of metrics $m$ (where $m \geq 1$) collected for a microservice, $t$ is the failure onset time, and $E$ is the failure resolution time. The algorithm uses the $\psi\text{-PC}$ tool to identify interventional targets and a Chi-squared ($\chi^2$) test to assess variable independence. To capture interventional effects, $\psi\text{-PC}$ introduces an auxiliary binary indicator node, $F-{\text{Node}} \in \{0, 1\}$, representing normal and anomalous operating modes, respectively. 
To optimize execution time, RCD avoids learning the complete causal graph and instead downsamples the input space. It randomly divides the dataset into smaller metric subsets of size $\gamma$ before executing $\psi\text{-PC}$. This process operates hierarchically: candidate interventional targets identified at each level are aggregated and passed to the subsequent level until the remaining subset is smaller than $\gamma$. While evaluated using the Sock-shop microservice benchmark with seven application-layer metrics---specifically CPU utilization, memory usage, system load, error rate (non-200 responses), and latencies ($P_{50}$, $P_{90}$, $P_{99}$)---the RCD algorithm \cite{rcd} is sound but not complete. 

Despite its design efficiencies, RCD exhibits two critical limitations: \begin{enumerate} \item \textbf{Algorithmic Inconsistency:} Randomly partitioning the input dataset introduces significant variance in the identified root causes across different executions, degrading localization accuracy on large datasets. \item \textbf{Layer-Insularity:} Restricting analysis to a few application-layer metrics ignores the broader cloud stack, reducing the diagnostic explainability required for automated remediation. \end{enumerate} 
To address these gaps, this study introduces Topology-Aware RCD (TA-RCD). A detailed breakdown of the TA-RCD framework and its constituent algorithms is presented in the following section.
\vspace{-0.3cm}
\subsection{Topology-Aware RCD (TA-RCD)}
\label{tarcd}
The proposed Topology-Aware RCD (TA-RCD) extends the baseline RCD framework by incorporating end-to-end service topology information to capture potential fault propagation paths across the cloud stack. By constraining root-cause discovery using topology-derived dependencies, TA-RCD improves localization accuracy while introducing minimal computational overhead. The root-cause localization problem for TA-RCD is formally defined as follows:

\begin{definitiongray}{Problem Statement}
For a performance anomaly reported at time $t_n$, let $D^*$ denote the anomalous metric observations collected over the look-back window $[t_n-d,t_n]$, where $d$ is the anomaly analysis duration. Let $D$ denote the normal metric observations collected over a known or configurable baseline window. Given $D$, $D^*$, and a dynamic service topology $G_T=(V_T,E_T)$, where metrics $\mathcal{M}$ are collected from cloud components represented in $V_T$, identify a ranked set of root-cause metrics $\mathcal{R}\subseteq\mathcal{M}$ that explains the observed deviation.
\end{definitiongray}

TA-RCD differs from RCD in its subset creation approach. The creation of subsets on which the  $\psi$-PC algorithm is applied in TA-RCD is based on the following observations:\\ 
\vspace{-0.9em}
\begin{enumerate}
    \item A fault changes the node's generative mechanism, and learning a complete causal graph is not required to identify the root cause \cite{rcd}.
    \item Faults, failures, and errors propagate across the components in a cloud stack \cite{drm_ucc}. This propagation induces a high noise-to-signal ratio in the data across the layers \cite{rcd}, lowering the accuracy of root cause identification techniques.
    \item Considering the end-to-end metrics for a cloud application service without acknowledging its layers and characteristics will reduce the accuracy of root cause localization techniques \cite{drm_ucc}.
    \item A fault or failure propagates across the components only if a link or path exists among the components in the service topology information. Factoring in this connectedness while forming subsets for executing $\psi$-PC can improve the accuracy of the root cause localization \cite{drm_topo}. 
    \item If a node is the origin of a fault, there will be an increase in the number of metrics from that given node affected by the fault or failure \cite{ediag}.
\end{enumerate}

Based on the above observations, instead of a random split of the input dataset as in RCD, TA-RCD performs a topology-aware split using the dynamic end-to-end service topology. Topology-aware grouping improves the causal discovery by (i) constraining the edges to be tested for CI, and (2) enabling the identification of failure correlations based on causal effects, while explaining the service behavior. As in RCD, TA-RCD also splits the dataset further into subsets using the configurable parameter $\gamma$. Further, $\gamma$ is set to 4 for all the experiments presented in this paper.

TA-RCD (Algorithm \ref{ta-rcd-alg}) operates in three phases to maximize localization accuracy and diagnostic explainability, evaluating four resource dimensions (CPU, memory, disk, and network) per component. 

\begin{enumerate}
    \item \textbf{Intra-Layer Phase:} Identifies anomalous metrics within each architectural layer across all resource types.
    \item \textbf{Topology-Aware Phase:} Inspects the end-to-end execution paths of the service topology to identify the origin of anomaly. It computes a weighted $p$-value based on the structural heuristic that the root-cause node exhibits the highest density of anomalous metrics.
    \item \textbf{Ranking Phase:} Aggregates and sorts the metric $p$-values generated from the conditional independence (CI) tests to prioritize the most likely root causes.
\end{enumerate}
The soundness of TA-RCD is directly inherited from the soundness of the baseline RCD algorithm \cite{rcd_sup}.

The Initial Phase (Steps 4 to 13 of Algorithm \ref{ta-rcd-alg}) partitions the normal ($D$) and anomalous ($D^*$) datasets into intra-layer subsets using layer and resource information based on service topology metadata \cite{drm_topo}. To optimize execution time, each layer- and resource-specific partition is downsampled into smaller subsets bounded by $\gamma$. Grouping metrics by resource type allows for localized comparisons, reducing false-positive causal links. The $\psi\text{-PC}$ algorithm executes hierarchically over these subsets, iteratively aggregating candidate root causes until the candidate pool size falls below $\gamma$.

The second phase (Steps 14 to 28 of Algorithm \ref{ta-rcd-alg}) inspects the fault propagation across the end-to-end execution paths in the topology, spanning from application service instances down to the physical hosts. These paths are pre-computed and cached, updated only upon a topology change signal. TA-RCD intersects the candidate root causes from the initial phase with these path dependencies, subdividing them by resource type into subsets of size $\gamma$ for hierarchical $\psi\text{-PC}$ execution. To prioritize metrics along the primary fault-propagation path, the conditional independence (CI) test $p$-values are scaled by the architectural path weight. This weight reflects the density of anomalies on path $P_i$ relative to the global anomaly count, defined as:

 \begin{equation} p_{ij}^{\text{weighted}} = p_{ij} \times \left( \frac{n_i}{N} \right) \end{equation} 

where $p_{ij}$ is the raw CI test $p$-value for metric $j$ on path $P_i$, $n_i$ is the count of suspected anomalous metrics on path $P_i$, and $N$ is the total count of suspected anomalous metrics across all paths. This weighting structure quantifies the statistical significance of Phase 1 candidates, filtering out spurious cross-path associations.

The third phase of TA-RCD (Step 30 of Algorithm \ref{ta-rcd-alg}) takes as input the reduced subset of the candidate root causes from the second phase and sorts them on the p-value of CI tests. 

The Theoretical Justification for the proposed Two-Phase Topology-Aware Grouping is as follows:

Let $G_T=(V_T,E_T)$ denote the dynamic end-to-end service topology of cloud components, and $G_C=(V_C,E_C)$ the latent causal graph over the metric space ($n=|V_C|$). We define a structural mapping function $c: V_C \to V_T \times \mathcal{L}$, where $\mathcal{L} = \{\text{Infrastructure}, \text{Platform}, \text{Application}\}$ denotes the architectural layers. TA-RCD isolates root causes via a two-phase topological mechanism: \textit{Phase 1} groups metrics per individual service layers, and \textit{Phase 2} traces anomalies along the end-to-end execution paths of $G_T$.

\begin{assumption}[Hierarchical Fault Propagation]
If a metric $X_i$ directly influences another metric $X_j$ during an anomaly, their structural coordinates must satisfy either:
\begin{enumerate}
    \item \textbf{Intra-Layer Alignment (Phase 1):} $v_i = v_j$ across layers $l_i \neq l_j$ (per-resource).
    \item \textbf{End-to-end Alignment (Phase 2):} $\mathrm{dist}_{G_T}(v_i, v_j) \le r$ for a bounded radius $r \ge 1$.
\end{enumerate}
\end{assumption}


Let $\mathcal{N}_{\text{local}}(X)$ denote the valid metric search space for metric $X$ under TA-RCD's phase-specific topology constraints, where $\mathcal{N}_{\text{local}}(X)$ contains only those candidate metrics that are topologically admissible for conditional independence testing with $X$. Let $k_{\text{local}} = |\mathcal{N}_{\text{local}}(X)|$ be the size of this local search space, with $k_{\text{local}} \ll n$, where $n$ is the total number of available metrics. Let $\mathrm{Pa}(X)$ and $\mathrm{Ch}(X)$ denote the parents and children of $X$ in the underlying causal graph, respectively, and let $\mathrm{Adj}(X) = \mathrm{Pa}(X) \cup \mathrm{Ch}(X)$ denote the immediate causal neighborhood of $X$. By Assumption 1, the true immediate causal neighborhood is topology-bounded such that $\mathrm{Adj}(X) \subseteq \mathcal{N}_{\text{local}}(X)$.
\begin{theorem}
Let $X$ be a true root cause and $Y \in \mathrm{Adj}(X)$. If $1 < \gamma \le k_{\text{local}} < n$, the two-phase topology-aware grouping yields a strictly higher probability of co-assigning $X$ and $Y$ to an execution subset of size $\gamma$ than stochastic random grouping.
\end{theorem}
\vspace{-0.2cm}
\begin{proof}
Constraint-based causal discovery requires $X$ and $Y$ to co-occur within the same partitioned calculation pool. Considering the following cases:

\noindent\textbf{Case 1: Random Grouping ($S_R$).} 
Sampling $\gamma$ metrics uniformly at random from the global space $V_C$ yields a hypergeometric distribution. Given $X$ is selected, the probability of including neighbor $Y$ is:
\begin{equation}
P(Y \in S_R \mid X \in S_R) = \frac{\binom{1}{1}\binom{n-2}{\gamma-2}}{\binom{n-1}{\gamma-1}} = \frac{\gamma-1}{n-1}
\end{equation}

\noindent\textbf{Case 2: Two-Phase Topology-Aware Grouping ($S_T$).}
TA-RCD strictly constrains the sampling domain to the phase workspace $\mathcal{N}_{\text{local}}(X)$, where $k_{\text{local}} = k_{\text{layer}}$ in Phase 1 and $k_{\text{local}} = k_{\text{path}}$ in Phase 2. Because $Y \in \mathrm{Adj}(X) \subseteq \mathcal{N}_{\text{local}}(X)$, the remaining $\gamma-1$ slots are chosen strictly from the local neighborhood pool:
\begin{equation}
P(Y \in S_T \mid X \in S_T) = \frac{\binom{1}{1}\binom{k_{\text{local}}-2}{\gamma-2}}{\binom{k_{\text{local}}-1}{\gamma-1}} = \frac{\gamma-1}{k_{\text{local}}-1}
\end{equation}

\noindent Given that a cloud deployment's global metric space scales heavily relative to localized component layers or immediate execution paths ($k_{\text{local}} \ll n$), it follows that:
\begin{equation}
k_{\text{local}} < n \implies \frac{\gamma-1}{k_{\text{local}}-1} > \frac{\gamma-1}{n-1}
\end{equation}
\begin{equation}
\text{Hence, } \quad P(Y \in S_T \mid X \in S_T) > P(Y \in S_R \mid X \in S_R)
\end{equation}
\end{proof}
Consequently, decoupling optimization into an intra-layer phase followed by an end-to-end path tracking phase maximizes causal coverage. This eliminates redundant conditional independence tests between disjoint architecture domains, drastically reducing the family-wise false-positive rate.
Thus, under topology-local fault propagation, TA-RCD improves causal-neighborhood coverage in both its layer-level and end-to-end phases. The topology is not treated as the true causal graph; rather, it is used as a structural prior to reduce unnecessary CI tests among topologically unrelated metrics and to focus causal discovery on feasible fault propagation paths.

\begin{algorithm}
\caption{Topology Aware-Root Cause Discovery (TA-RCD)}\label{alg:cap}
\label{ta-rcd-alg}
\scriptsize
\hspace*{\algorithmicindent}\textbf{Input} {Normal dataset D, Anomalous dataset D$\ast$, $\gamma$, $\psi$-PC \cite{psi_pc}, k : Max. no. of root causes, Topology information $\textit{td}$}\\
\hspace*{\algorithmicindent}\textbf{Output} {A list of k root causes, RC}      
\begin{algorithmic}[1]
\Procedure{TA-RCD}{D, D$\ast$, $\gamma$, \textit{td}}
    \State STOPO $\gets$ service topology information from td .json file \cite{drm_topo}
    \State K $\gets$ Set of metrics of D, $D\ast$
    \medskip
    \State  \textbf{$\#$Phase-1}
    \While{$|K| > k $}
        \State $S_1$ $\gets$ Layer-resource specific partitioning of $|K|$ into subsets of size $\gamma$ for all layers in STOPO
        \State R $\gets$ 0
        \For{\textbf{all} $s \in S_1$}
            \State G $\gets$ $\psi$-PC(D[s],$D^\ast$[s])
            \State R $\gets$ R $\cup$ $Ne_G$(F-NODE)
        \EndFor
        \State K $\gets$ R
    \EndWhile
    \medskip
    \State \textbf{$\#$Phase-2: Topology-Aware Phase}
    \State $RC_2$ $\gets$ 0
    \For{\textbf{all} end-to-end paths EP in STOPO}
        \State $w_p$ = $|$metrics from K in R$|$ / $|$K$|$
        \State RC\_EP $\gets$ Metrics across EP in $RC_2$
        \State $S_2$ $\gets$ resource specific partitioning of RC\_EP into subsets of size $\gamma$
        \For{\textbf{all} $s \in S_2$}
            \State G $\gets$ $\psi$-PC(D[s],$D^\ast$[s])
            \State R $\gets$ R $\cup$ $Ne_G$(F-NODE)
            \For{$r \in R$}
                \State Weighted p-value $\gets$ p-value $* w_p$
            \EndFor
        \EndFor
        \State $RC_2$ $\gets$ $RC_2$ $\cup$ R
    \EndFor
    \medskip
    \State  \textbf{$\#$Phase-3: Ranking Phase}
    \State $RC$ $\gets$ Sort $RC_2$ based on the p-value of CI tests
\EndProcedure
\State \Return{} $RC$;
\end{algorithmic}
\end{algorithm}

\begin{figure}[!htb]
    \centering  \includegraphics[width=\linewidth,keepaspectratio]{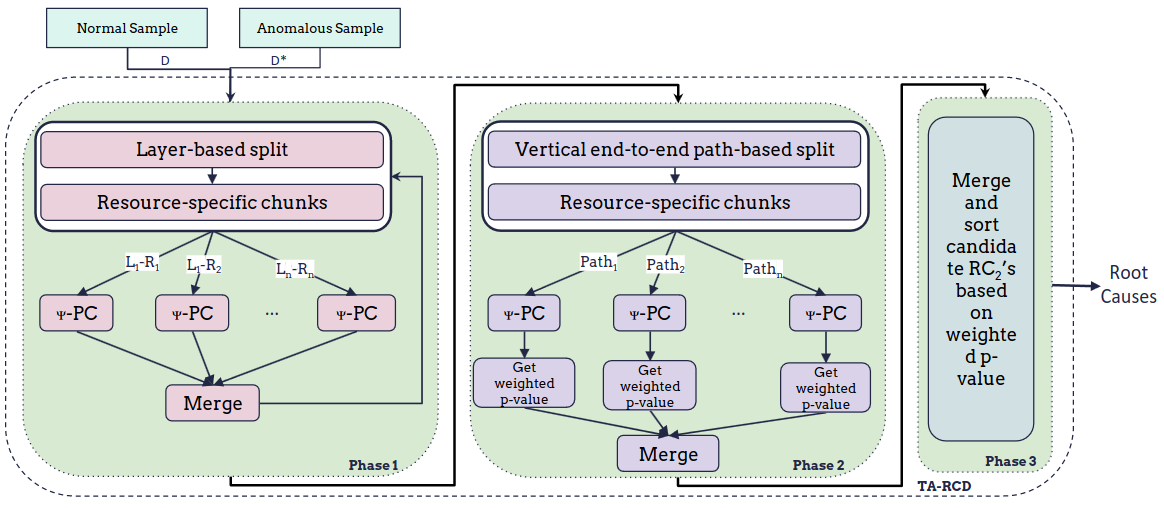}
    \caption{TA-RCD Framework}\label{fig:tarcd}
\end{figure}

The TA-RCD framework is depicted in Fig. \ref{fig:tarcd}.

\textit{Time and Space Complexity Analysis: }
Let $m$ denote the number of metrics used for root-cause localization. Both RCD and TA-RCD avoid learning a complete causal graph, whose complexity can be exponential, and instead learn localized causal relations by performing Conditional Independence (CI) tests involving the $F$-Node. Assuming each CI test has $O(1)$ complexity, RCD progressively reduces the metric set by a factor of $\gamma$ across hierarchical iterations, resulting in an overall time complexity of $O(m)$.

TA-RCD extends this process by applying localized causal discovery across $l$ cloud service layers and $r$ resource groups, such as CPU, memory, disk, and network. Its total time complexity is $O(lrm) + O(lr) + O(lr\log(lr))$. Since the number of layers and resource groups is small and bounded in practice, TA-RCD remains effectively linear in the number of metrics while improving localization through topology- and resource-aware partitioning.

The space complexity of RCD is $O(m)$. TA-RCD additionally stores the cross-layer service topology as an adjacency list, resulting in space complexity $O(m) + O(n+e)$, where $n$ and $e$ denote the number of topology nodes and edges, respectively.

\section{Evaluation}
\label{eval}

\vspace{-0.6cm}

\subsection{Experimental Design} 
The evaluation addresses three primary research questions: 
\begin{enumerate}
\item RQ1: How does TA-RCD compare with existing root-cause analysis baselines?
\item RQ2: Does topology awareness improve the efficiency of root-cause discovery?
\item RQ3: Does the proposed metric selection enhance explainability in root-cause localization?
\end{enumerate}

\vspace{-0.5cm}
\subsection{Evaluation metrics}
Following established literature, performance is evaluated using two standard metrics: Accuracy at $k$ ($AC@k$) and Average Accuracy at $k$ ($Avg@k$). Given a set of failure cases $\mathcal{A}$, $AC@k$ measures the probability that the true root causes are included within the top-$k$ recommendations:\begin{equation}AC@k = \frac{1}{|\mathcal{A}|} \sum_{a \in \mathcal{A}} \frac{\sum_{i=1}^{k} \mathbb{I}\left(R_a[i] \in V_a^{rc}\right)}{\min(k, |V_a^{rc}|)}\end{equation}\noindent where $R_a[i]$ is the $i$-th ranked metric recommended for failure case $a$, $V_a^{rc}$ is the ground-truth root-cause set, and $\mathbb{I}(\cdot)$ is the indicator function. $AC@k$ values range from 0 to 1, with higher values indicating better localized accuracy.To reflect overall ranking quality across all granularities up to $k$, $Avg@k$ is calculated as:
\vspace{-0.5cm}
\begin{equation}Avg@k = \frac{1}{k} \sum_{j=1}^{k} AC@j\end{equation}

\subsection{RQ1: Comparison with Existing RCA Baselines}
To address RQ1, TA-RCD is evaluated on the open-source RCAEval benchmark~\cite{rcaeval}. RCAEval provides a standardized framework containing 735 real failure cases with annotated ground truths (root-cause services and indicators) across three benchmark microservice systems: Online Boutique, Sock-Shop, and Train-Ticket. Further, RCAEval records the state signals across the end-to-ens service components. The dataset covers resource faults (CPU, memory, disk, and socket stress) injected via stress-ng, and network faults (delays and packet drops) injected via tc, with corresponding root-cause indicators mapped to resource usage and network performance metrics.

TA-RCD is compared against three metric-based baselines available in RCAEval: CIRCA~\cite{causal_infer}, RCD~\cite{rcd}, and BARO~\cite{baro}. These baselines are selected because they also isolate root causes by analyzing monitoring metrics or metric-derived causal and statistical relationships, making them directly comparable to TA-RCD. Other baselines, namely RUN~\cite{run} and CausalRCA~\cite{causalrca}, were excluded from the final evaluation because their executions failed to complete within the predefined timeout budget (2hrs in this case). Table~\ref{tab:rcaeval_re2_datasets} outlines the dataset characteristics.

\begin{table}[t]
\centering
\caption{RE2 datasets from RCAEval used for comparison with TA-RCD}
\label{tab:rcaeval_re2_datasets}
\scriptsize
\setlength{\tabcolsep}{4pt}
\begin{tabular}{l l c c c c}
\hline
\textbf{Dataset} & \textbf{Application} & \textbf{\#Services} & \textbf{\#Cases} & \textbf{Fault Types} & \textbf{\#Metrics} \\
\hline
RE2-OB & Online Boutique & 12 & 90 & All RE2 faults\textsuperscript{*} & 72--461 \\
RE2-SS & Sock Shop & 15 & 90 & All RE2 faults\textsuperscript{*} & 1509--1581 \\
RE2-TT & Train Ticket & 64 & 90 & All RE2 faults\textsuperscript{*} & 433--452 \\
\hline
\multicolumn{6}{l}{\textsuperscript{*}CPU, MEM, DISK, DELAY, LOSS, and SOCKET.}
\end{tabular}
\end{table}
\subsubsection{Topology Construction} 
For the RCAEval datasets, the structural topology is constructed from deployment metadata for each benchmark run. Cluster-level container information and pod-to-node placement configurations are mapped to generate a unified, cross-layer topology representation for each application. Specifically, this mapping binds each logical service to its corresponding container, physical pod, and hosting virtual machine or node.

\subsubsection{Evaluation Metrics and Scores}
To evaluate localization accuracy, results are reported using three standard ranking metrics: AC@3, AC@5, and Avg@5.

Each ranking metric is reported as an ordered pair $(m,s)$, where $m$ and $s$ denote metric-level and service-level localization accuracy, respectively. Service-level accuracy checks whether the faulty service is identified, while metric-level accuracy strictly checks whether the predicted fault-indicating metric category matches the ground truth. Since actionable remediation requires identifying the exact cause, such as CPU, memory, disk, delay, loss, or socket faults, the better approach is determined primarily by its ability to localize the correct metric category, not just the affected service.

\subsubsection{Results}
Table~\ref{tab:rq1_rcaeval_results} compares TA-RCD with metric-based RCA baselines on the RCAEval RE2 datasets. Each score is reported as $(m,s)$ for metric-level and service-level localization. The green entries indicate the best result for a given metric, fault type, and dataset, while the red entries indicate the worst result.
\begin{table*}[t]
\centering
\caption{Comparison of root-cause localization performance}
\label{tab:rq1_rcaeval_results}
\setlength{\tabcolsep}{1.2pt} 
\renewcommand{\arraystretch}{1.35} 
\footnotesize 

\resizebox{\textwidth}{!}{
\begin{tabular}{ll ccc ccc ccc ccc ccc ccc c}
\hline
\textbf{Dataset} & \textbf{Method} & \multicolumn{3}{c}{\textbf{CPU}} & \multicolumn{3}{c}{\textbf{MEM}} & \multicolumn{3}{c}{\textbf{DISK}} & \multicolumn{3}{c}{\textbf{SOCKET}} & \multicolumn{3}{c}{\textbf{DELAY}} & \multicolumn{3}{c}{\textbf{LOSS}} & \textbf{Avg. Exec. Time} \\
 & & AC@3 & AC@5 & Avg@5 & AC@3 & AC@5 & Avg@5 & AC@3 & AC@5 & Avg@5 & AC@3 & AC@5 & Avg@5 & AC@3 & AC@5 & Avg@5 & AC@3 & AC@5 & Avg@5 & \textbf{(sec/case)} \\
\hline
\hline

\multirow{4}{*}{\parbox{2cm}{Online Boutique\\(RE2-OB)}} 
 & CIRCA & (0.25,0.25) & (0.62,0.62) & (0.26,0.26) & \textcolor{red}{(0,0.20)} & \textcolor{red}{(0,0.60)} & \textcolor{red}{(0,0.24)} & \textcolor{red}{(0,0.20)} & \textcolor{red}{(0,0.60)} & \textcolor{red}{(0,0.24)} & \textcolor{red}{(0,0.20)} & \textcolor{red}{(0,0.60)} & \textcolor{red}{(0,0.24)} & \textcolor{red}{(0,0.20)} & \textcolor{red}{(0,0.60)} & \textcolor{red}{(0,0.24)} & \textcolor{red}{(0,0.20)} & \textcolor{red}{(0,0.60)} & \textcolor{red}{(0,0.24)} & 0.17 \\
 & RCD & \textcolor{red}{(0,1.00)} & \textcolor{red}{(0,1.00)} & \textcolor{red}{(0,1.00)} & (0.20,0.60) & (0.27,0.67) & (0.19,0.60) & (0,0.67) & (0,0.67) & (0,0.65) & (0,0.80) & (0,0.80) & (0,0.79) & \textcolor{darkgreen}{(0.60,0.60)} & (0.60,0.60) & \textcolor{darkgreen}{(0.60,0.60)} & (0.07,0.40) & (0.07,0.47) & (0.07,0.40) & 7.60 \\
 & BARO & (0.06,0.25) & (0.06,0.50) & (0.05,0.24) & (0.47,0.67) & (0.53,0.87) & (0.45,0.68) & (0,0.33) & (0,0.80) & (0,0.45) & (0,0.27) & (0,0.73) & (0,0.31) & (0.27,0.27) & (0.53,0.60) & (0.24,0.28) & \textcolor{darkgreen}{(0.67,0.80)} & \textcolor{darkgreen}{(0.67,0.93)} & \textcolor{darkgreen}{(0.56,0.73)} & 0.33 \\
 & TA-RCD & \textcolor{darkgreen}{(0.31,0.94)} & \textcolor{darkgreen}{(0.62,1.00)} & \textcolor{darkgreen}{(0.40,0.88)} & \textcolor{darkgreen}{(0.60,0.80)} & \textcolor{darkgreen}{(0.73,0.93)} & \textcolor{darkgreen}{(0.63,0.80)} & \textcolor{darkgreen}{(0,1.00)} & \textcolor{darkgreen}{(0,1.00)} & \textcolor{darkgreen}{(0,0.84)} & \textcolor{darkgreen}{(0,0.93)} & \textcolor{darkgreen}{(0,1.00)} & \textcolor{darkgreen}{(0,0.80)} & (0.47,0.67) & \textcolor{darkgreen}{(0.60,0.87)} & (0.41,0.63) & (0.47,0.67) & (0.6,0.87) & (0.44,0.69) & 14.34 \\
\hline

\multirow{4}{*}{\parbox{2cm}{Train Ticket\\(RE2-TT)}} 
 & CIRCA & \textcolor{red}{(0,0)} & \textcolor{red}{(0,0)} & \textcolor{red}{(0,0)} & \textcolor{red}{(0,0)} & \textcolor{red}{(0,0)} & \textcolor{red}{(0,0)} & \textcolor{red}{(0,0)} & \textcolor{red}{(0,0)} & \textcolor{red}{(0,0)} & \textcolor{red}{(0,0)} & \textcolor{red}{(0,0)} & \textcolor{red}{(0,0)} & \textcolor{red}{(0,0)} & \textcolor{red}{(0,0)} & \textcolor{red}{(0,0)} & \textcolor{red}{(0,0)} & \textcolor{red}{(0,0)} & \textcolor{red}{(0,0)} & 0.69 \\
 & RCD & (0,0.53) & (0,0.53) & (0,0.51) & (0.07,0.20) & (0.13,0.27) & (0.08,0.17) & (0,0.33) & (0,0.33) & (0,0.29) & \textcolor{darkgreen}{(0.20,0.33)} & \textcolor{darkgreen}{(0.20,0.33)} &\textcolor{darkgreen}{(0.17,0.29)} & (0,0.07) & (0,0.07) & (0,0.07) & \textcolor{red}{(0,0)} & \textcolor{red}{(0,0)} & \textcolor{red}{(0,0)} & 16.78 \\
 & BARO & (0,0.07) & (0,0.13) & (0,0.08) & (0.07,0.13) & (0.13,0.33) & (0.07,0.17) & \textcolor{darkgreen}{(0.87,1.00)} & \textcolor{darkgreen}{(1.00,1.00)} & \textcolor{darkgreen}{(0.77,0.96)} & \textcolor{red}{(0,0)} & \textcolor{red}{(0,0)} & \textcolor{red}{(0,0)} & \textcolor{darkgreen}{(0.13,0.20)} & \textcolor{darkgreen}{(0.20,0.27)} & \textcolor{darkgreen}{(0.12,0.17)} & (0.07,0.07) & (0.05,0.07) & (0.04,0.05) & 1.36 \\
 & TA-RCD & \textcolor{darkgreen}{(0.72,0.76)} & \textcolor{darkgreen}{(0.72,0.89)} & \textcolor{darkgreen}{(0.66,0.79)} & \textcolor{darkgreen}{(0.22,0.69)} & \textcolor{darkgreen}{(0.27,0.82)} & \textcolor{darkgreen}{(0.23,0.70)} & (0.53,0.67) & (0.60,0.80) & (0.53,0.68) & (0,0.56) & (0,0.71) & (0,0.57) & (0.07,0.31) & (0.07,0.49) & (0.07,0.33) & \textcolor{darkgreen}{(0.07,0.38)} & \textcolor{darkgreen}{(0.09,0.56)} & \textcolor{darkgreen}{(0.0,0.37)} & 28.42 \\
\hline

\multirow{4}{*}{\parbox{2cm}{Sock Shop\\(RE2-SS)}} 
 & CIRCA & \textcolor{darkgreen}{(0.40,0.40)} & (0.40,0.80) & (0.32,0.48) & \textcolor{red}{(0,0.40)} & \textcolor{red}{(0,0.80)} & \textcolor{red}{(0,0.48)} & \textcolor{red}{(0,0.40)} & \textcolor{red}{(0,0.80)} & \textcolor{red}{(0,0.48)} & \textcolor{red}{(0,0.40)} & \textcolor{red}{(0,0.80)} & \textcolor{red}{(0,0.48)} & \textcolor{red}{(0,0.40)} & \textcolor{red}{(0,0.80)} & \textcolor{red}{(0,0.48)} & \textcolor{red}{(0,0.40)} & \textcolor{red}{(0,0.80)} & \textcolor{red}{(0,0.48)} & 0.17 \\
 & RCD & (0.07,0.80) & \textcolor{red}{(0.07,0.87)} & (0.07,0.79) & (0.20,0.53) & (0.20,0.53) & (0.17,0.48) & \textcolor{darkgreen}{(0,0.73)} & (0,0.73) & (0,0.71) & (0,0.87) & (0,0.93) & (0,0.88) & (0.27,0.60) & (0.27,0.60) & (0.24,0.52) & (0.07,0.27) & (0.07,0.33) & (0.07,0.24) & 4.60 \\
 & BARO & \textcolor{red}{(0,0.33)} & (0.20,0.87) & \textcolor{red}{(0.05,0.39)} & \textcolor{darkgreen}{(0.73,0.73)} & (0.73,1.00) & \textcolor{darkgreen}{(0.69,0.80)} & (0,0.60) & (0,0.93) & (0,0.52) & (0,0.47) & \textcolor{red}{(0,0.67)} & \textcolor{red}{(0,0.44)} & (0.20,0.27) & \textcolor{darkgreen}{(0.47,0.73)} & \textcolor{darkgreen}{(0.27,0.37)} & \textcolor{darkgreen}{(0.60,0.73)} & \textcolor{darkgreen}{(0.73,1.00)} & \textcolor{darkgreen}{(0.59,0.77)} & 0.22 \\
 & TA-RCD & (0.33,0.73) & \textcolor{darkgreen}{(0.47,1.00)} & \textcolor{darkgreen}{(0.32,0.72)} & (0.67,0.87) & \textcolor{darkgreen}{(0.80,1.00)} & (0.60,0.80) & \textcolor{darkgreen}{(0,0.73)} & \textcolor{darkgreen}{(0,1.00)} & \textcolor{darkgreen}{(0,0.72)} & \textcolor{darkgreen}{(0.07,0.93)} & \textcolor{darkgreen}{(0.07,1.00)} & \textcolor{darkgreen}{(0.07,0.88)} & \textcolor{darkgreen}{(0.33,0.53)} & (0.47,0.67) & (0.25,0.43) & (0.33,0.73) & (0.67,0.93) & (0.39,0.64) & 16.01 \\
\hline
\end{tabular}%
}
\end{table*}
\subsubsection{Discussion}
Overall, TA-RCD provides the most consistent root-cause localization performance across the RE2 benchmark. Based on the metric component $m$ of the reported $(m,s)$ pairs, TA-RCD improves average AC@5 by 0.06 over the strongest baseline, corresponding to an 18\% relative improvement, and improves Avg@5 by 0.04, corresponding to a 19\% relative improvement. These gains show that TA-RCD more reliably ranks the explanatory metric category among the top candidates, which is important for reliable remediation decisions. Across Online Boutique, Sock Shop, and the more complex Train Ticket topology, TA-RCD achieves the best or near-best results for most fault categories and improves service-level localization.

Although RCD performed better in an individual category, this behavior is inconsistent because randomized metric grouping may either preserve or separate causally related metrics. BARO also outperforms TA-RCD in selected categories, which can be attributed to its use of Multivariate Bayesian Online Change Point Detection. By modeling dependencies within multivariate time-series metrics, BARO can accurately detect abrupt distributional changes when a fault produces a sharp, localized anomaly within a correlated metric group. However, BARO does not explicitly incorporate deployment topology, and therefore, its advantage is less consistent when the observed anomaly includes propagated effects across service, runtime, and infrastructure layers. In contrast, TA-RCD preserves topology-consistent relationships across these layers, yielding more robust and stable localization accuracy across datasets and fault types.

The average execution time of TA-RCD is higher than CIRCA, BARO, and RCD because it performs topology-aware partitioning, cross-layer candidate construction, and end-to-end causal refinement. However, the runtime increase is bounded by the structured grouping of metrics and does not require exhaustive causal discovery over the full metric space. The measured execution time remains practical for real-time or near-real-time anomaly localization, where root-cause analysis is triggered after anomaly detection. When considered together with its higher localization accuracy, TA-RCD provides a favorable trade-off between execution time, diagnostic precision, and explainability.
\vspace{-1.0em}

\subsection{RQ2: Topology Awareness for Efficient Root-Cause Discovery}
To evaluate the impact of topology awareness on root-cause discovery, we compare RCD and TA-RCD using controlled fault-injection experiments on a locally deployed Sock-Shop microservice application \cite{sockshop}.

\subsubsection{Experimental Setup}
Following RCD \cite{rcd}, the application was deployed on a Kubernetes cluster comprising one master and 15 worker nodes, each running 16 KVM-based virtual machines. Each worker VM was configured with 2 vCPUs and 2\,GB of RAM, hosted on a physical server equipped with 2 2.10 GHz Intel Xeon processors (12 hyperthreaded cores each) and 128 \,GB of RAM.

Application workload was generated using Locust \cite{locust}. To emulate realistic user behavior, Selenium \cite{selenium} was integrated with Locust to ensure user requests followed browser-level workflows rather than purely protocol-level HTTP calls. Each simulated user executed a standardized transaction sequence consisting of authentication, product browsing, and order placement. The request volume followed a normal distribution with a mean of 50 users per second, matching the baseline RCD evaluation environment.

Telemetry metrics across the cloud stack were collected via Prometheus \cite{prom}. Service-level metrics were gathered using the monitoring manifests provided by Sock-Shop. Container-level metrics were aggregated using cAdvisor \cite{cadvisor}, VM-level KVM metrics via libvirt-exporter \cite{libvirt_exporter}, and physical or virtual resource metrics using Node Exporter \cite{node_exporter} with the \texttt{perf}, \texttt{processes}, and \texttt{network\_route} collectors enabled.

Faults were injected into both the container and VM layers to evaluate cross-layer fault propagation. Representing common resource-saturation anomalies in cloud environments, the evaluation considered CPU hog, memory leak, and network delay faults. CPU and memory faults were injected via stress-ng \cite{stress-ng}, whereas network delays were introduced using the \texttt{tc} tool \cite{tc_tool}. Anomalies were targeted at the \texttt{cart}, \texttt{catalogue}, \texttt{orders}, \texttt{payment}, and \texttt{user} microservices, which constitute the critical user-facing paths in Sock-Shop \cite{rcd}. For the ground-truth root causes, CPU utilization was designated as the root-cause metric for CPU hog; memory utilization for memory leak; and network transmission throughput for network delay.

Each experimental execution consisted of a 5-minute steady-state interval under normal operating conditions for baseline telemetry collection, followed by a 5-minute post-injection interval to capture anomalous behavior. Each localization algorithm was evaluated over 100 repeated runs for each dataset. Performance is reported using AC@3 and AC@5.

\subsubsection{Results and Discussion}
This experiment isolates the effect of topology awareness by comparing RCD and TA-RCD under identical metric sets and injected-fault conditions. While RCD serves as the topology-agnostic baseline, TA-RCD uses the known end-to-end topology to guide candidate selection and causal refinement. The comparison therefore evaluates whether incorporating topology improves root-cause localization beyond standard metric-space causal discovery.

\begin{figure*}[!htb]
   \centering
   \begin{minipage}{0.16\textwidth}
     \centering
     \includegraphics[width=\linewidth]{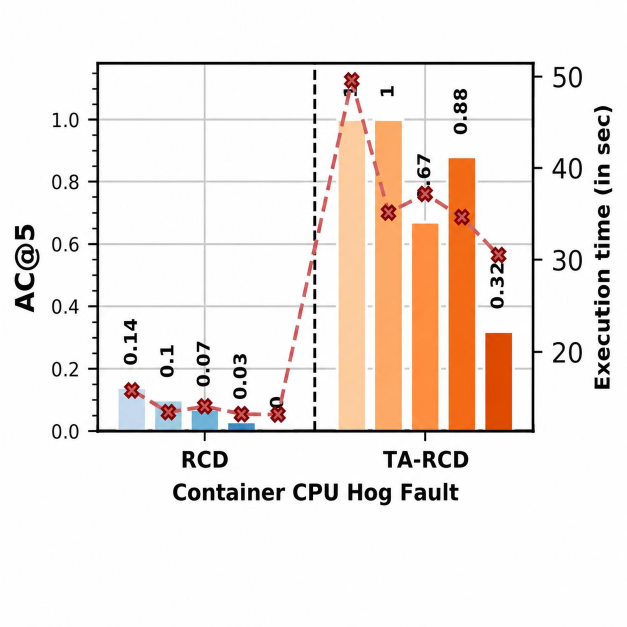}

     {\small (a)}
   \end{minipage}
   \hfill
   \begin{minipage}{0.16\textwidth}
     \centering
     \includegraphics[width=\linewidth]{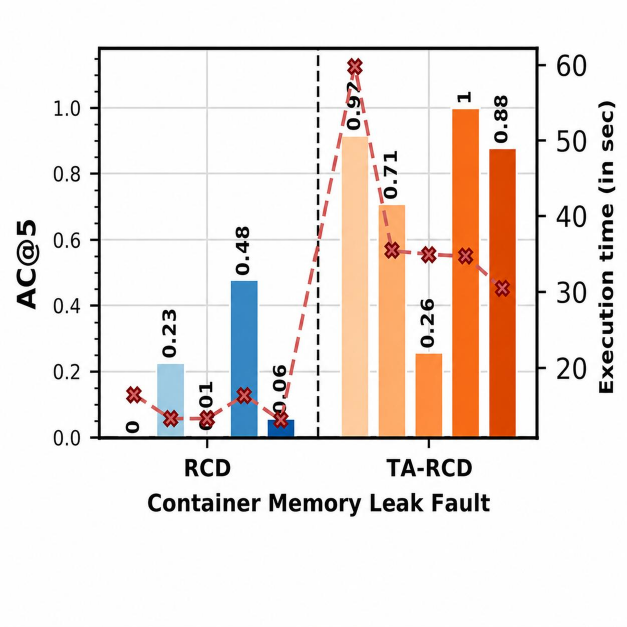}

     {\small (b)}
   \end{minipage}
   \hfill
   \begin{minipage}{0.16\textwidth}
     \centering
     \includegraphics[width=\linewidth]{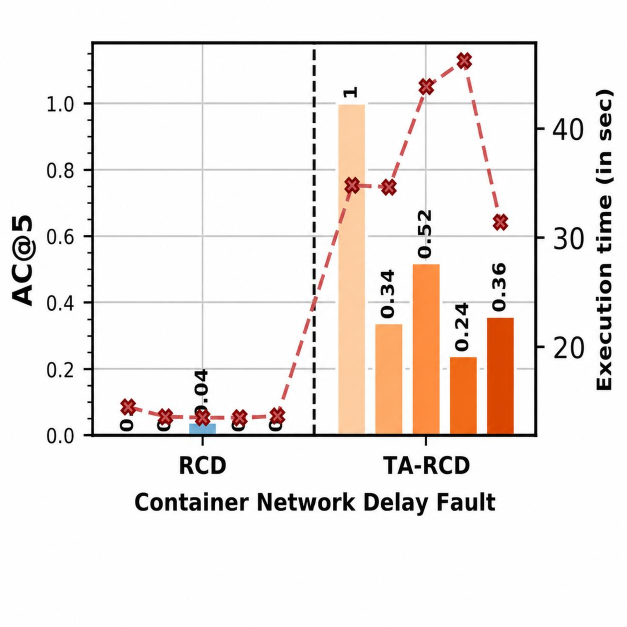}

     {\small (c)}
   \end{minipage}
   \hfill
   \begin{minipage}{0.16\textwidth}
     \centering
     \includegraphics[width=\linewidth]{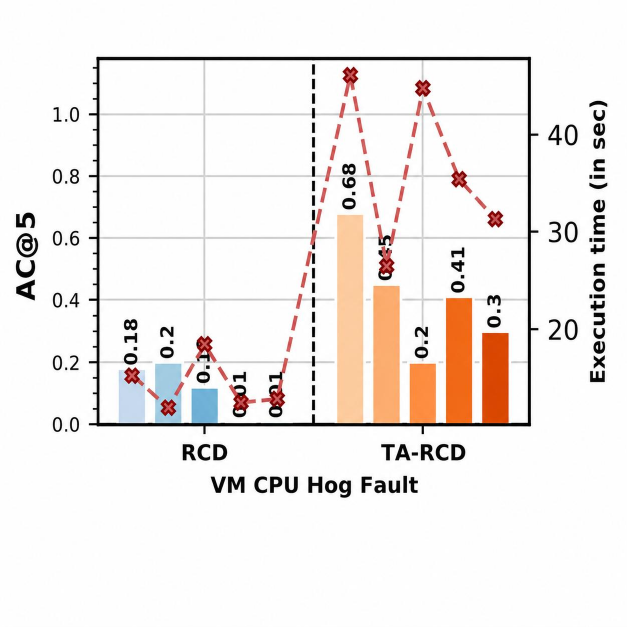}

     {\small (d)}
   \end{minipage}
   \hfill
   \begin{minipage}{0.16\textwidth}
     \centering
     \includegraphics[width=\linewidth]{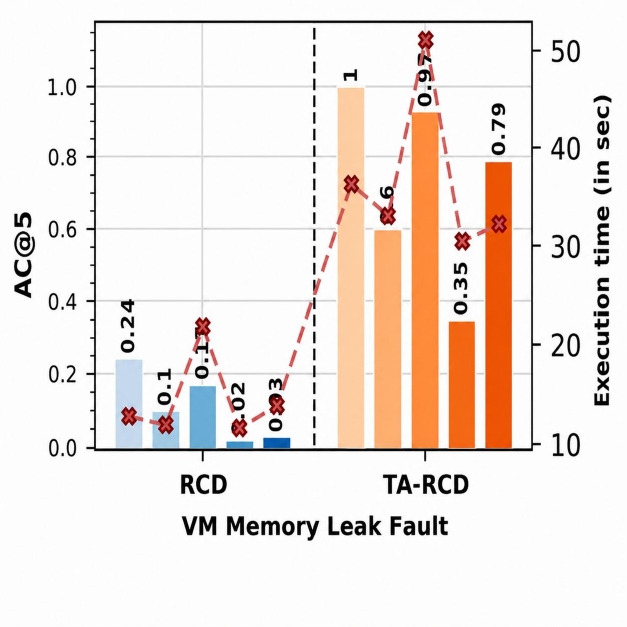}

     {\small (e)}
   \end{minipage}
   \hfill
   \begin{minipage}{0.16\textwidth}
     \centering
     \includegraphics[width=\linewidth]{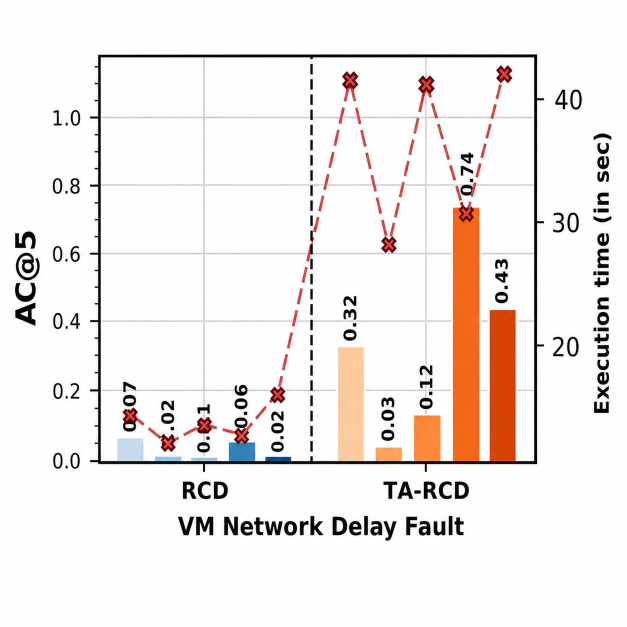}

     {\small (f)}
   \end{minipage}

   \vspace{1mm}

   \begin{minipage}{0.49\textwidth}
     \centering
     \includegraphics[width=\linewidth]{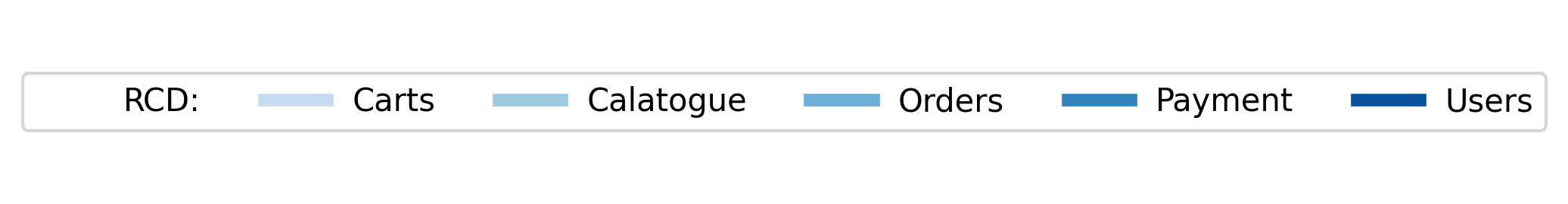}
   \end{minipage}
   \hfill
   \begin{minipage}{0.49\textwidth}
     \centering
     \includegraphics[width=\linewidth]{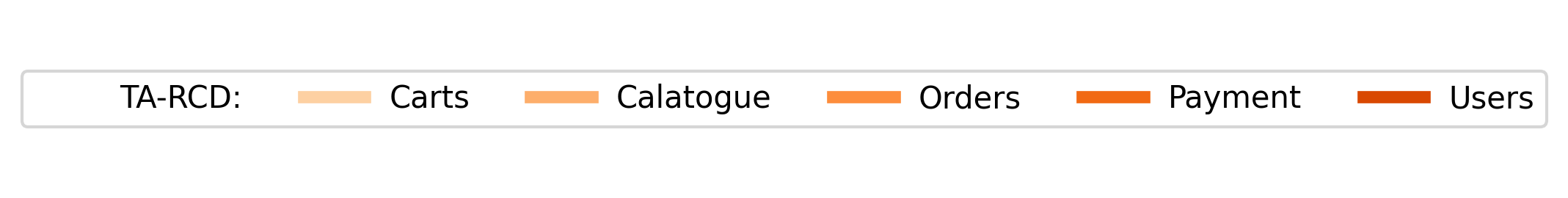}
   \end{minipage}

   \vspace{-0.5cm}
   \caption{Difference in AC@5 for anomalies injected on Sock-shop microservice application}
   \label{fig:top5}
\end{figure*}

\begin{figure*}[!htb]
   \centering
   \begin{minipage}{0.16\textwidth}
     \centering
     \includegraphics[width=\linewidth]{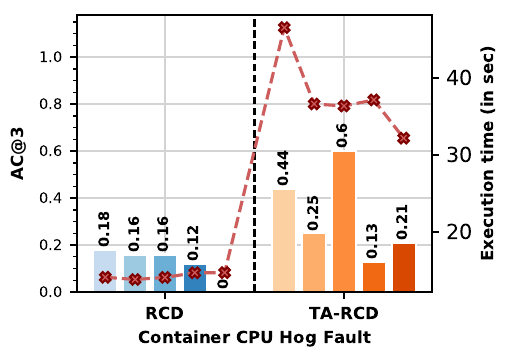}

     {\small (a)}
   \end{minipage}
   \hfill
   \begin{minipage}{0.16\textwidth}
     \centering
     \includegraphics[width=\linewidth]{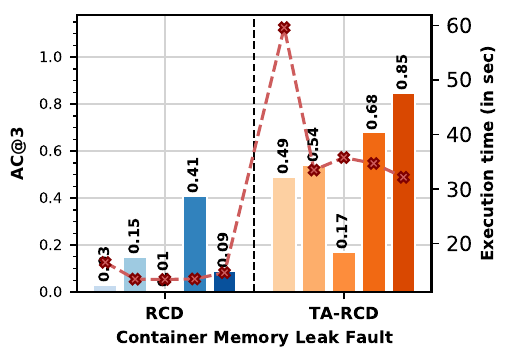}

     {\small (b)}
   \end{minipage}
   \hfill
   \begin{minipage}{0.16\textwidth}
     \centering
     \includegraphics[width=\linewidth]{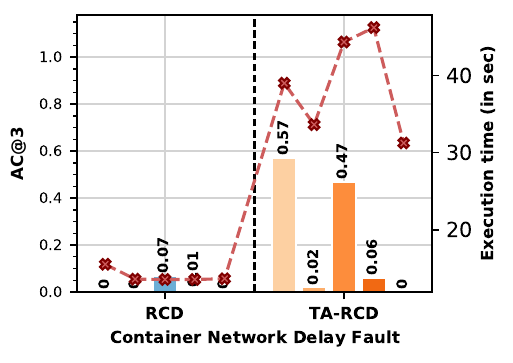}

     {\small (c)}
   \end{minipage}
   \hfill
   \begin{minipage}{0.16\textwidth}
     \centering
     \includegraphics[width=\linewidth]{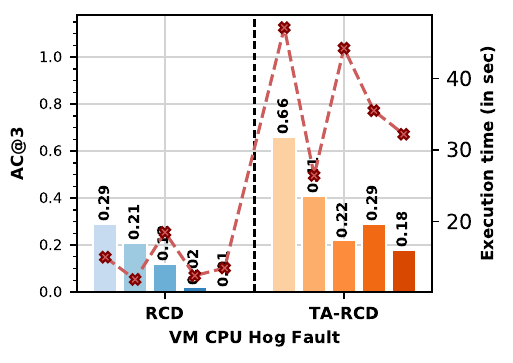}

     {\small (d)}
   \end{minipage}
   \hfill
   \begin{minipage}{0.16\textwidth}
     \centering
     \includegraphics[width=\linewidth]{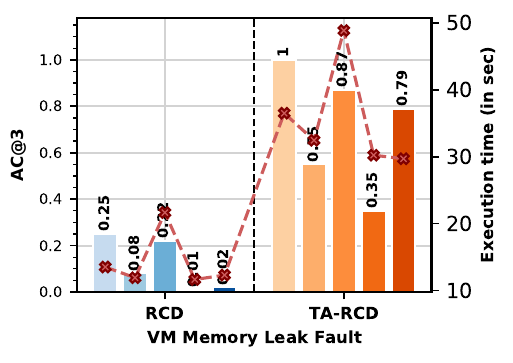}

     {\small (e)}
   \end{minipage}
   \hfill
   \begin{minipage}{0.16\textwidth}
     \centering
     \includegraphics[width=\linewidth]{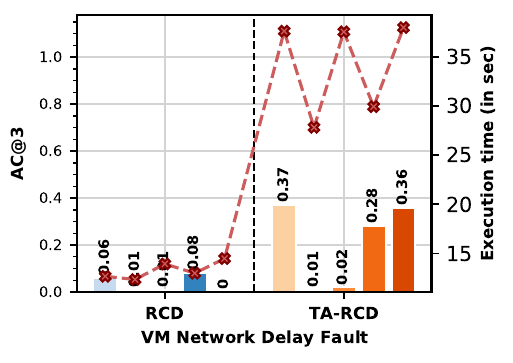}

     {\small (f)}
   \end{minipage}

   \vspace{1mm}

   \begin{minipage}{0.49\textwidth}
     \centering
     \includegraphics[width=\linewidth]{figures/legend1_new.png}
   \end{minipage}
   \hfill
   \begin{minipage}{0.49\textwidth}
     \centering
     \includegraphics[width=\linewidth]{figures/legend2_new.png}
   \end{minipage}

   \vspace{-0.5cm}
   \caption{Difference in AC@3 for anomalies injected on Sock-shop microservice application}
   \label{fig:top3}
\end{figure*}

The results show that topology awareness substantially improves localization accuracy. For a CPU hog fault injected into the \texttt{orders} container, RCD identified the ground-truth container CPU utilization metric in only 7 out of 100 runs, whereas TA-RCD localized it in 67 runs. This improvement is due to TA-RCD’s ability to first narrow anomalous candidates within each layer and then refine them using cross-layer service-topology paths. Across the scenarios in Fig.~\ref{fig:top5}(a)--(f), TA-RCD improves AC@5 by 2$\times$--21$\times$ over RCD, and Fig.~\ref{fig:top3}(a)--(f) shows a 4$\times$--13$\times$ improvement for AC@3. These results demonstrate that topology-aware grouping enables more consistent and reliable root-cause discovery across repeated trials.

As observed earlier, TA-RCD introduces additional runtime due to topology-guided partitioning and localized causal refinement.
Furthermore, the partition-based grouping operations within TA-RCD are inherently parallelizable, offering a clear trajectory for future optimizations to reduce runtime latency.

\vspace{-0.2cm}
\subsection{RQ3: Does the proposed metric selection enhance explainability in root-cause localization?}
To evaluate diagnostic explainability, the root-cause outputs of RCD and TA-RCD are analyzed under a CPU-hog fault injected into the \texttt{orders} container, where the target root cause is its CPU utilization metric. For this, the study uses both state and activity metrics.

\subsubsection{Explainability of TA-RCD}
\begin{figure}[!htb]
   \begin{minipage}{1.0\linewidth}
     \centering
     \includegraphics[width=1.0\linewidth]{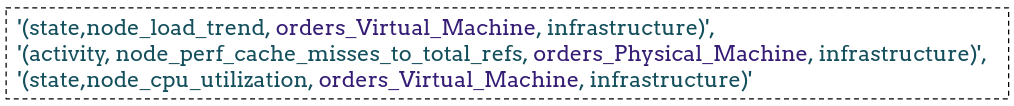}
     \caption{Top-5 fault identified by RCD for a container CPU hog fault in orders microservice} 
     \label{fig:res_rcd}
   \end{minipage}
   \vspace{0.00mm}
   \begin{minipage}{1\linewidth}
     \centering
     \includegraphics[width=1.0\linewidth]{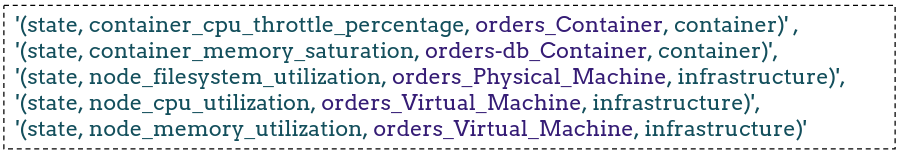}
     \caption{Top-5 fault identified by TA-RCD for a container CPU hog fault in orders microservice}
     \label{fig:res_tarcd}
   \end{minipage}
   \vspace{0.00mm}
\end{figure}

Fig.~\ref{fig:res_rcd} and Fig.~\ref{fig:res_tarcd} present representative Top-5 root-cause sets for RCD (which identifies only three metrics) and TA-RCD, respectively. As shown in Fig.~\ref{fig:res_tarcd}, TA-RCD isolates the CPU saturation signal at the \texttt{orders} container and explicitly maps it to the hosting physical node's CPU utilization. This cross-layer association structurally contexts the fault, tracing the anomaly from the container layer to the infrastructure host. This topology-consistent propagation path makes the output highly actionable for automated remediation.

In contrast, RCD's output (Fig.~\ref{fig:res_rcd}) lacks deployment context. Because RCD evaluates metric groups independently of service architecture, it often localizes structurally disconnected metrics. Even when capturing anomalies, it fails to elucidate how they causally relate to the injected fault, limiting its utility for explainable diagnosis.

\subsubsection{Effect of Topology-Aware Grouping}
To isolate the impact of architectural awareness, RCD outputs were analyzed across failed and successful localization runs under the same \texttt{orders} container CPU-hog scenario. 

\begin{figure}[!h]
\includegraphics[width=\linewidth,keepaspectratio]{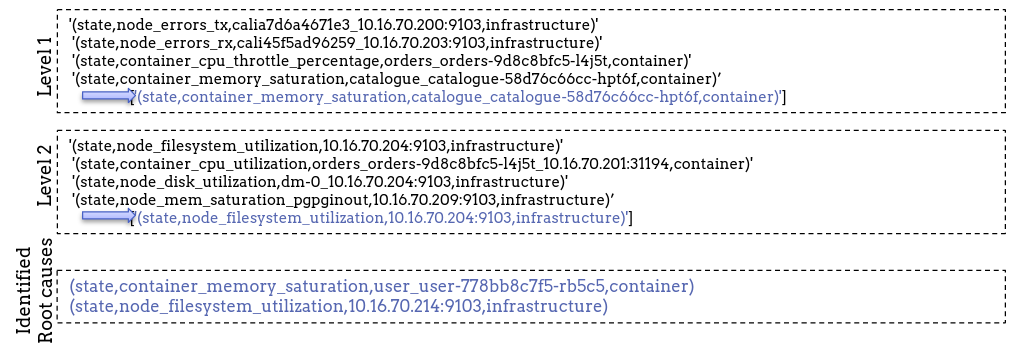}
\caption{Groups formed in RCD during an unsuccessful root-cause localization} 
\label{fig:unsuccessful}
\end{figure}

\begin{figure}[!h]
\includegraphics[width=\linewidth,keepaspectratio]{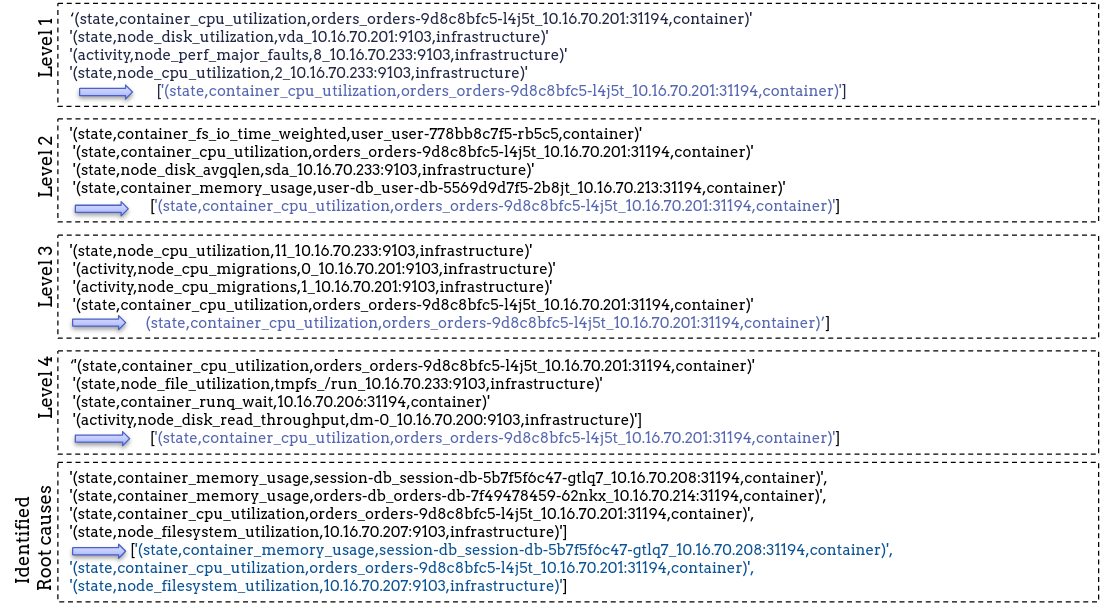}
\caption{Groups formed in RCD during a successful root-cause localization} 
\label{fig:successful}
\end{figure}

Fig.~\ref{fig:unsuccessful} illustrates a failed RCD execution where the target metric, \textit{(state, container\_cpu\_utilization, orders\_orders-9d8c8bfc5-l4j5t\_10.16.70.201:31194, container)}, is prematurely eliminated at Level 2. This occurs because RCD's randomized grouping can separate topologically dependent metrics or aggregate entirely unrelated ones, eroding the causal integrity of the discovery space. Conversely, Fig.~\ref{fig:successful} highlights a successful RCD run where randomized grouping happenstance aligned with the underlying topology: \texttt{10.16.70.201} maps to the VM hosting the \texttt{orders} service, and \texttt{10.16.70.233} corresponds to the physical host. While RCD can succeed when random partitions accidentally preserve structural relationships, this behavior is non-deterministic.

TA-RCD addresses this limitation by anchoring its grouping process within the infrastructure topology. It systematically partitions metrics by resource type, cloud layer, and execution path, ensuring structurally dependent metrics are co-retained to generate topology-consistent explanations.

\subsubsection{Discussion}
The evaluation indicates that the proposed metric selection and topology-aware analysis enhance explainability by embedding outputs within the multi-layer context of resource saturation and deployment structure. By correlating service-level anomalies with container, VM, and host telemetry, TA-RCD delivers superior localization accuracy and more actionable diagnostic context than topology-agnostic baselines.

\section{Conclusion}
\label{conclusion}
This work introduced TA-RCD, a topology-aware framework for root-cause localization that provides explainable failure diagnosis in cloud-native environments. By leveraging deployment topology, service dependencies, and cross-layer telemetry, TA-RCD isolates both the anomalous component and the exact fault-indicating metric necessary for actionable remediation. Experimental evaluation demonstrates significant performance gains over state-of-the-art methods. In a controlled testbed environment, TA-RCD improves average AC@5 by at least 2$\times$ and AC@3 by at least 4$\times$ compared to standard RCD. Validation on the public RCAEval benchmark further confirms its effectiveness, with approximately 18\% improvement in metric-level AC@5 and 19\% improvement in Avg@5 over the strongest baseline among CIRCA, RCD, and BARO. Future work will focus on reducing runtime latency by parallelizing the partition-level causal discovery operations.


\begin{thebibliography}{10}
\providecommand{\url}[1]{#1}
\csname url@samestyle\endcsname
\providecommand{\newblock}{\relax}
\providecommand{\bibinfo}[2]{#2}
\providecommand{\BIBentrySTDinterwordspacing}{\spaceskip=0pt\relax}
\providecommand{\BIBentryALTinterwordstretchfactor}{4}
\providecommand{\BIBentryALTinterwordspacing}{\spaceskip=\fontdimen2\font plus
\BIBentryALTinterwordstretchfactor\fontdimen3\font minus \fontdimen4\font\relax}
\providecommand{\BIBforeignlanguage}[2]{{%
\expandafter\ifx\csname l@#1\endcsname\relax
\typeout{** WARNING: IEEEtran.bst: No hyphenation pattern has been}%
\typeout{** loaded for the language `#1'. Using the pattern for}%
\typeout{** the default language instead.}%
\else
\language=\csname l@#1\endcsname
\fi
#2}}
\providecommand{\BIBdecl}{\relax}
\BIBdecl

\bibitem{cloud_computing}
\BIBentryALTinterwordspacing
M.~Armbrust, A.~Fox, R.~Griffith, A.~D. Joseph, R.~Katz, A.~Konwinski, G.~Lee, D.~Patterson, A.~Rabkin, I.~Stoica, and M.~Zaharia, ``A view of cloud computing,'' \emph{Commun. ACM}, vol.~53, no.~4, p. 50–58, apr 2010. [Online]. Available: \url{https://doi.org/10.1145/1721654.1721672}
\BIBentrySTDinterwordspacing

\bibitem{drm_ucc}
\BIBentryALTinterwordspacing
D.~R. Mathews, M.~Verma, P.~Aggarwal, and J.~Lakshmi, ``Towards failure correlation for improved cloud application service resilience,'' in \emph{Proceedings of the 14th IEEE/ACM International Conference on Utility and Cloud Computing Companion}, ser. UCC '21.\hskip 1em plus 0.5em minus 0.4em\relax New York, NY, USA: Association for Computing Machinery, 2022. [Online]. Available: \url{https://doi.org/10.1145/3492323.3495586}
\BIBentrySTDinterwordspacing

\bibitem{res_defn}
L.~Simoncini, ``Resilient computing: An engineering discipline,'' in \emph{2009 IEEE International Symposium on Parallel \& Distributed Processing}, 2009, pp. 1--1.

\bibitem{drm1}
\BIBentryALTinterwordspacing
D.~R. Mathews and J.~Lakshmi, ``Service resilience framework for enhanced end-to-end service quality,'' in \emph{Proceedings of the 18th Workshop on Adaptive and Reflexive Middleware}, ser. ARM '19.\hskip 1em plus 0.5em minus 0.4em\relax New York, NY, USA: Association for Computing Machinery, 2019, p. 7–12. [Online]. Available: \url{https://doi.org/10.1145/3366612.3368123}
\BIBentrySTDinterwordspacing

\bibitem{rcd}
\BIBentryALTinterwordspacing
M.~A. Ikram, S.~Chakraborty, S.~Mitra, S.~Saini, S.~Bagchi, and M.~Kocaoglu, ``Root cause analysis of failures in microservices through causal discovery,'' in \emph{Advances in Neural Information Processing Systems}, A.~H. Oh, A.~Agarwal, D.~Belgrave, and K.~Cho, Eds., 2022. [Online]. Available: \url{https://openreview.net/forum?id=weoLjoYFvXY}
\BIBentrySTDinterwordspacing

\bibitem{deathstar}
\BIBentryALTinterwordspacing
Y.~Gan, Y.~Zhang, D.~Cheng, A.~Shetty, P.~Rathi, N.~Katarki, A.~Bruno, J.~Hu, B.~Ritchken, B.~Jackson, K.~Hu, M.~Pancholi, Y.~He, B.~Clancy, C.~Colen, F.~Wen, C.~Leung, S.~Wang, L.~Zaruvinsky, M.~Espinosa, R.~Lin, Z.~Liu, J.~Padilla, and C.~Delimitrou, ``An open-source benchmark suite for microservices and their hardware-software implications for cloud \& edge systems,'' in \emph{Proceedings of the Twenty-Fourth International Conference on Architectural Support for Programming Languages and Operating Systems}, ser. ASPLOS '19.\hskip 1em plus 0.5em minus 0.4em\relax New York, NY, USA: Association for Computing Machinery, 2019, p. 3–18. [Online]. Available: \url{https://doi.org/10.1145/3297858.3304013}
\BIBentrySTDinterwordspacing

\bibitem{dso}
C.~Sridharan, \emph{Distributed Systems Observability}, 2018.

\bibitem{metric_defn}
C.~Sauvanaud, K.~Lazri, M.~Kaâniche, and K.~Kanoun, ``Anomaly detection and root cause localization in virtual network functions,'' in \emph{2016 IEEE 27th International Symposium on Software Reliability Engineering (ISSRE)}, 2016, pp. 196--206.

\bibitem{metric_expln}
\BIBentryALTinterwordspacing
L.~Wu, J.~Bogatinovski, S.~Nedelkoski, J.~Tordsson, and O.~Kao, ``Performance diagnosis in cloud microservices using deep learning,'' in \emph{Service-Oriented Computing – ICSOC 2020 Workshops: AIOps, CFTIC, STRAPS, AI-PA, AI-IOTS, and Satellite Events, Dubai, United Arab Emirates, December 14–17, 2020, Proceedings}.\hskip 1em plus 0.5em minus 0.4em\relax Berlin, Heidelberg: Springer-Verlag, 2020, p. 85–96. [Online]. Available: \url{https://doi.org/10.1007/978-3-030-76352-7_13}
\BIBentrySTDinterwordspacing

\bibitem{sieve}
\BIBentryALTinterwordspacing
J.~Thalheim, A.~Rodrigues, I.~E. Akkus, P.~Bhatotia, R.~Chen, B.~Viswanath, L.~Jiao, and C.~Fetzer, ``Sieve: Actionable insights from monitored metrics in distributed systems,'' in \emph{Proceedings of the 18th ACM/IFIP/USENIX Middleware Conference}, ser. Middleware '17.\hskip 1em plus 0.5em minus 0.4em\relax New York, NY, USA: Association for Computing Machinery, 2017, p. 14–27. [Online]. Available: \url{https://doi.org/10.1145/3135974.3135977}
\BIBentrySTDinterwordspacing

\bibitem{ubl}
\BIBentryALTinterwordspacing
D.~J. Dean, H.~Nguyen, and X.~Gu, ``Ubl: Unsupervised behavior learning for predicting performance anomalies in virtualized cloud systems,'' in \emph{Proceedings of the 9th International Conference on Autonomic Computing}, ser. ICAC '12.\hskip 1em plus 0.5em minus 0.4em\relax New York, NY, USA: Association for Computing Machinery, 2012, p. 191–200. [Online]. Available: \url{https://doi.org/10.1145/2371536.2371572}
\BIBentrySTDinterwordspacing

\bibitem{drm_topo}
D.~R. Mathews, M.~Verma, J.~Lakshmi, and P.~Aggarwal, ``Towards more effective and explainable fault management using cross-layer service topology,'' in \emph{2022 IEEE 15th International Conference on Cloud Computing (CLOUD)}, 2022, pp. 94--96.

\bibitem{sre_golden_sig}
``Monitoring distributed systems,'' \url{https://sre.google/sre-book/monitoring-distributed-systems/}, (Accessed on 05/08/2023).

\bibitem{bgregg}
B.~Gregg, \emph{Systems Performance: Enterprise and the Cloud}, 1st~ed.\hskip 1em plus 0.5em minus 0.4em\relax USA: Prentice Hall Press, 2013.

\bibitem{red}
``Red method,'' \url{https://www.weave.works/blog/the-red-method-key-metrics-for-microservices-architecture/}, (Accessed on 05/08/2023).

\bibitem{microcause}
Y.~Meng, S.~Zhang, Y.~Sun, R.~Zhang, Z.~Hu, Y.~Zhang, C.~Jia, Z.~Wang, and D.~Pei, ``Localizing failure root causes in a microservice through causality inference,'' in \emph{2020 IEEE/ACM 28th International Symposium on Quality of Service (IWQoS)}, 2020, pp. 1--10.

\bibitem{ad_container}
Q.~Du, T.~Xie, and Y.~He, ``Anomaly detection and diagnosis for container-based microservices with performance monitoring,'' in \emph{Algorithms and Architectures for Parallel Processing}, J.~Vaidya and J.~Li, Eds.\hskip 1em plus 0.5em minus 0.4em\relax Cham: Springer International Publishing, 2018, pp. 560--572.

\bibitem{frlmpfg}
\BIBentryALTinterwordspacing
Y.~Chen, D.~Xu, N.~Chen, and X.~Wu, ``Frl-mfpg: Propagation-aware fault root cause location for microservice intelligent operation and maintenance,'' \emph{Information and Software Technology}, vol. 153, p. 107083, 2023. [Online]. Available: \url{https://www.sciencedirect.com/science/article/pii/S0950584922001926}
\BIBentrySTDinterwordspacing

\bibitem{perf_diag_micro}
\BIBentryALTinterwordspacing
L.~Wu, J.~Bogatinovski, S.~Nedelkoski, J.~Tordsson, and O.~Kao, ``Performance diagnosis in cloud microservices using deep learning,'' in \emph{Service-Oriented Computing – ICSOC 2020 Workshops: AIOps, CFTIC, STRAPS, AI-PA, AI-IOTS, and Satellite Events, Dubai, United Arab Emirates, December 14–17, 2020, Proceedings}.\hskip 1em plus 0.5em minus 0.4em\relax Berlin, Heidelberg: Springer-Verlag, 2020, p. 85–96. [Online]. Available: \url{https://doi.org/10.1007/978-3-030-76352-7_13}
\BIBentrySTDinterwordspacing

\bibitem{microrca}
L.~Wu, J.~Tordsson, E.~Elmroth, and O.~Kao, ``Microrca: Root cause localization of performance issues in microservices,'' in \emph{NOMS 2020 - 2020 IEEE/IFIP Network Operations and Management Symposium}, 2020, pp. 1--9.

\bibitem{fd4c}
T.~Wang, W.~Zhang, C.~Ye, J.~Wei, H.~Zhong, and T.~Huang, ``Fd4c: Automatic fault diagnosis framework for web applications in cloud computing,'' \emph{IEEE Transactions on Systems, Man, and Cybernetics: Systems}, vol.~46, no.~1, pp. 61--75, 2016.

\bibitem{fchain}
H.~Nguyen, Z.~Shen, Y.~Tan, and X.~Gu, ``Fchain: Toward black-box online fault localization for cloud systems,'' in \emph{2013 IEEE 33rd International Conference on Distributed Computing Systems}, 2013, pp. 21--30.

\bibitem{loud}
L.~Mariani, C.~Monni, M.~Pezzé, O.~Riganelli, and R.~Xin, ``Localizing faults in cloud systems,'' in \emph{2018 IEEE 11th International Conference on Software Testing, Verification and Validation (ICST)}, 2018, pp. 262--273.

\bibitem{wang2023hierarchical}
D.~Wang, Z.~Chen, J.~Ni, L.~Tong, Z.~Wang, Y.~Fu, and H.~Chen, ``Hierarchical graph neural networks for causal discovery and root cause localization,'' 2023.

\bibitem{automap}
\BIBentryALTinterwordspacing
M.~Ma, J.~Xu, Y.~Wang, P.~Chen, Z.~Zhang, and P.~Wang, ``Automap: Diagnose your microservice-based web applications automatically,'' in \emph{Proceedings of The Web Conference 2020}, ser. WWW '20.\hskip 1em plus 0.5em minus 0.4em\relax New York, NY, USA: Association for Computing Machinery, 2020, p. 246–258. [Online]. Available: \url{https://doi.org/10.1145/3366423.3380111}
\BIBentrySTDinterwordspacing

\bibitem{cloudranger}
\BIBentryALTinterwordspacing
P.~Wang, J.~Xu, M.~Ma, W.~Lin, D.~Pan, Y.~Wang, and P.~Chen, ``Cloudranger: Root cause identification for cloud native systems,'' in \emph{Proceedings of the 18th IEEE/ACM International Symposium on Cluster, Cloud and Grid Computing}, ser. CCGrid '18.\hskip 1em plus 0.5em minus 0.4em\relax IEEE Press, 2018, p. 492–502. [Online]. Available: \url{https://doi.org/10.1109/CCGRID.2018.00076}
\BIBentrySTDinterwordspacing

\bibitem{causeinfer}
P.~Chen, Y.~Qi, P.~Zheng, and D.~Hou, ``Causeinfer: Automatic and distributed performance diagnosis with hierarchical causality graph in large distributed systems,'' in \emph{IEEE INFOCOM 2014 - IEEE Conference on Computer Communications}, 2014, pp. 1887--1895.

\bibitem{rca_pc}
J.~Weng, J.~H. Wang, J.~Yang, and Y.~Yang, ``Root cause analysis of anomalies of multitier services in public clouds,'' \emph{IEEE/ACM Transactions on Networking}, vol.~26, no.~4, pp. 1646--1659, 2018.

\bibitem{cid_smallmet}
I.~Cohen, M.~Goldszmidt, T.~Kelly, J.~Symons, and J.~S. Chase, ``Correlating instrumentation data to system states: A building block for automated diagnosis and control,'' in \emph{Proceedings of the 6th Conference on Symposium on Operating Systems Design \& Implementation - Volume 6}, ser. OSDI'04.\hskip 1em plus 0.5em minus 0.4em\relax USA: USENIX Association, 2004, p.~16.

\bibitem{hora}
T.~Pitakrat, D.~Okanovic, A.~Van~Hoorn, and L.~Grunske, ``An architecture-aware approach to hierarchical online failure prediction,'' in \emph{2016 12th International ACM SIGSOFT Conference on Quality of Software Architectures (QoSA)}, 2016, pp. 60--69.

\bibitem{hora1}
\BIBentryALTinterwordspacing
T.~Pitakrat, D.~Okanović, A.~{van Hoorn}, and L.~Grunske, ``Hora: Architecture-aware online failure prediction,'' \emph{Journal of Systems and Software}, vol. 137, pp. 669--685, 2018. [Online]. Available: \url{https://www.sciencedirect.com/science/article/pii/S0164121217300390}
\BIBentrySTDinterwordspacing

\bibitem{fail_pred1}
T.~Chalermarrewong, T.~Achalakul, and S.~C.~W. See, ``Failure prediction of data centers using time series and fault tree analysis,'' in \emph{2012 IEEE 18th International Conference on Parallel and Distributed Systems}, 2012, pp. 794--799.

\bibitem{prepare}
Y.~Tan, H.~Nguyen, Z.~Shen, X.~Gu, C.~Venkatramani, and D.~Rajan, ``Prepare: Predictive performance anomaly prevention for virtualized cloud systems,'' in \emph{2012 IEEE 32nd International Conference on Distributed Computing Systems}, 2012, pp. 285--294.

\bibitem{lep}
\BIBentryALTinterwordspacing
X.~Zhou, X.~Peng, T.~Xie, J.~Sun, C.~Ji, D.~Liu, Q.~Xiang, and C.~He, ``Latent error prediction and fault localization for microservice applications by learning from system trace logs,'' in \emph{Proceedings of the 2019 27th ACM Joint Meeting on European Software Engineering Conference and Symposium on the Foundations of Software Engineering}, ser. ESEC/FSE 2019.\hskip 1em plus 0.5em minus 0.4em\relax New York, NY, USA: Association for Computing Machinery, 2019, p. 683–694. [Online]. Available: \url{https://doi.org/10.1145/3338906.3338961}
\BIBentrySTDinterwordspacing

\bibitem{fl_backgnd}
\BIBentryALTinterwordspacing
M.~Steinder and A.~S. Sethi, ``The present and future of event correlation: A need for end-to-end service fault localization,'' 2001. [Online]. Available: \url{https://api.semanticscholar.org/CorpusID:8070393}
\BIBentrySTDinterwordspacing

\bibitem{caus_mining}
\BIBentryALTinterwordspacing
J.~Qiu, Q.~Du, K.~Yin, S.-L. Zhang, and C.~Qian, ``A causality mining and knowledge graph based method of root cause diagnosis for performance anomaly in cloud applications,'' \emph{Applied Sciences}, vol.~10, no.~6, 2020. [Online]. Available: \url{https://www.mdpi.com/2076-3417/10/6/2166}
\BIBentrySTDinterwordspacing

\bibitem{root_cause_detect_soa}
\BIBentryALTinterwordspacing
M.~Kim, R.~Sumbaly, and S.~Shah, ``Root cause detection in a service-oriented architecture,'' in \emph{Proceedings of the ACM SIGMETRICS/International Conference on Measurement and Modeling of Computer Systems}, ser. SIGMETRICS '13.\hskip 1em plus 0.5em minus 0.4em\relax New York, NY, USA: Association for Computing Machinery, 2013, p. 93–104. [Online]. Available: \url{https://doi.org/10.1145/2465529.2465753}
\BIBentrySTDinterwordspacing

\bibitem{baro}
\BIBentryALTinterwordspacing
L.~Pham, H.~Ha, and H.~Zhang, ``Baro: Robust root cause analysis for microservices via multivariate bayesian online change point detection,'' 2024. [Online]. Available: \url{https://arxiv.org/abs/2405.09330}
\BIBentrySTDinterwordspacing

\bibitem{causalrca}
\BIBentryALTinterwordspacing
R.~Xin, P.~Chen, and Z.~Zhao, ``Causalrca: Causal inference based precise fine-grained root cause localization for microservice applications,'' \emph{Journal of Systems and Software}, vol. 203, p. 111724, 2023. [Online]. Available: \url{http://dx.doi.org/10.1016/j.jss.2023.111724}
\BIBentrySTDinterwordspacing

\bibitem{torai}
\BIBentryALTinterwordspacing
L.~Pham, H.~Ha, X.~Zhang, and H.~Zhang, ``Torai: Multi-source root cause analysis for blind spots in microservice service call graph,'' 2026. [Online]. Available: \url{https://arxiv.org/abs/2604.13522}
\BIBentrySTDinterwordspacing

\bibitem{causal_infer}
\BIBentryALTinterwordspacing
M.~Li, Z.~Li, K.~Yin, X.~Nie, W.~Zhang, K.~Sui, and D.~Pei, ``Causal inference-based root cause analysis for online service systems with intervention recognition,'' in \emph{Proceedings of the 28th ACM SIGKDD Conference on Knowledge Discovery and Data Mining}, ser. KDD '22.\hskip 1em plus 0.5em minus 0.4em\relax New York, NY, USA: Association for Computing Machinery, 2022, p. 3230–3240. [Online]. Available: \url{https://doi.org/10.1145/3534678.3539041}
\BIBentrySTDinterwordspacing

\bibitem{sage}
\BIBentryALTinterwordspacing
Y.~Gan, M.~Liang, S.~Dev, D.~Lo, and C.~Delimitrou, ``Sage: Practical and scalable ml-driven performance debugging in microservices,'' in \emph{Proceedings of the 26th ACM International Conference on Architectural Support for Programming Languages and Operating Systems}, ser. ASPLOS '21.\hskip 1em plus 0.5em minus 0.4em\relax New York, NY, USA: Association for Computing Machinery, 2021, p. 135–151. [Online]. Available: \url{https://doi.org/10.1145/3445814.3446700}
\BIBentrySTDinterwordspacing

\bibitem{metric_rank}
C.~Wu, N.~Zhao, L.~Wang, X.~Yang, S.~Li, M.~Zhang, X.~Jin, X.~Wen, X.~Nie, W.~Zhang, K.~Sui, and D.~Pei, ``Identifying root-cause metrics for incident diagnosis in online service systems,'' in \emph{2021 IEEE 32nd International Symposium on Software Reliability Engineering (ISSRE)}, 2021, pp. 91--102.

\bibitem{perf_diag_dl}
\BIBentryALTinterwordspacing
L.~Wu, J.~Bogatinovski, S.~Nedelkoski, J.~Tordsson, and O.~Kao, ``Performance diagnosis in cloud microservices using deep learning,'' in \emph{Service-Oriented Computing – ICSOC 2020 Workshops: AIOps, CFTIC, STRAPS, AI-PA, AI-IOTS, and Satellite Events, Dubai, United Arab Emirates, December 14–17, 2020, Proceedings}.\hskip 1em plus 0.5em minus 0.4em\relax Berlin, Heidelberg: Springer-Verlag, 2020, p. 85–96. [Online]. Available: \url{https://doi.org/10.1007/978-3-030-76352-7_13}
\BIBentrySTDinterwordspacing

\bibitem{ediag}
\BIBentryALTinterwordspacing
H.~Shan, Y.~Chen, H.~Liu, Y.~Zhang, X.~Xiao, X.~He, M.~Li, and W.~Ding, ``??-diagnosis: Unsupervised and real-time diagnosis of small- window long-tail latency in large-scale microservice platforms,'' in \emph{The World Wide Web Conference}, ser. WWW '19.\hskip 1em plus 0.5em minus 0.4em\relax New York, NY, USA: Association for Computing Machinery, 2019, p. 3215–3222. [Online]. Available: \url{https://doi.org/10.1145/3308558.3313653}
\BIBentrySTDinterwordspacing

\bibitem{mrca}
\BIBentryALTinterwordspacing
Y.~Wang, Z.~Zhu, Q.~Fu, Y.~Ma, and P.~He, ``Mrca: Metric-level root cause analysis for microservices via multi-modal data,'' in \emph{Proceedings of the 39th IEEE/ACM International Conference on Automated Software Engineering}, ser. ASE '24.\hskip 1em plus 0.5em minus 0.4em\relax New York, NY, USA: Association for Computing Machinery, 2024, p. 1057–1068. [Online]. Available: \url{https://doi.org/10.1145/3691620.3695485}
\BIBentrySTDinterwordspacing

\bibitem{run}
\BIBentryALTinterwordspacing
C.-M. Lin, C.~Chang, W.-Y. Wang, K.-D. Wang, and W.-C. Peng, ``Root cause analysis in microservice using neural granger causal discovery,'' 2024. [Online]. Available: \url{https://arxiv.org/abs/2402.01140}
\BIBentrySTDinterwordspacing

\bibitem{rcd_sup}
``Rcd supplemental material,'' \url{https://proceedings.neurips.cc/paper_files/paper/2022/file/c9fcd02e6445c7dfbad6986abee53d0d-Supplemental-Conference.pdf}, (Accessed on 05/08/2023).

\bibitem{psi_pc}
\BIBentryALTinterwordspacing
A.~Jaber, M.~Kocaoglu, K.~Shanmugam, and E.~Bareinboim, ``Causal discovery from soft interventions with unknown targets: Characterization and learning,'' in \emph{Advances in Neural Information Processing Systems}, H.~Larochelle, M.~Ranzato, R.~Hadsell, M.~Balcan, and H.~Lin, Eds., vol.~33.\hskip 1em plus 0.5em minus 0.4em\relax Curran Associates, Inc., 2020, pp. 9551--9561. [Online]. Available: \url{https://proceedings.neurips.cc/paper_files/paper/2020/file/6cd9313ed34ef58bad3fdd504355e72c-Paper.pdf}
\BIBentrySTDinterwordspacing

\bibitem{rcaeval}
\BIBentryALTinterwordspacing
L.~Pham, H.~Zhang, H.~Ha, F.~Salim, and X.~Zhang, ``Rcaeval: A benchmark for root cause analysis of microservice systems with telemetry data,'' 2025. [Online]. Available: \url{https://arxiv.org/abs/2412.17015}
\BIBentrySTDinterwordspacing

\bibitem{sockshop}
``Sock shop microservice,'' \url{https://github.com/microservices-demo/microservices-demo}, (Accessed on 05/08/2023).

\bibitem{locust}
``Locust load test,'' \url{https://locust.io/}, (Accessed on 05/08/2023).

\bibitem{selenium}
S.~M. Shariff, H.~Li, C.-P. Bezemer, A.~E. Hassan, T.~H. Nguyen, and P.~Flora, ``Improving the testing efficiency of selenium-based load tests,'' in \emph{2019 IEEE/ACM 14th International Workshop on Automation of Software Test (AST)}, 2019, pp. 14--20.

\bibitem{prom}
``Prometheus monitoringe,'' \url{https://prometheus.io/}, (Accessed on 05/08/2023).

\bibitem{cadvisor}
``cadvisor,'' \url{https://github.com/google/cadvisor}, (Accessed on 05/08/2023).

\bibitem{libvirt_exporter}
``Libvirt exporter,'' \url{https://hub.docker.com/r/alekseizakharov/libvirt-exporter}, (Accessed on 05/08/2023).

\bibitem{node_exporter}
``Prometheus node exporter,'' \url{https://github.com/prometheus/node_exporter}, (Accessed on 05/08/2023).

\bibitem{stress-ng}
``Stress-ng tool,'' \url{https://wiki.ubuntu.com/Kernel/Reference/stress-ng}, (Accessed on 05/08/2023).

\bibitem{tc_tool}
``tc tool,'' \url{https://man7.org/linux/man-pages/man8/tc.8.html}, (Accessed on 05/08/2023).

\end{thebibliography}


\vfill

\end{document}